\documentclass[12pt,preprint]{aastex}

\slugcomment{To be published in ApJ, 2009} 
\shorttitle{Reddening and Distance of IC 10} 
\shortauthors{Kim et al.}
\begin{document}

\title{ Reddening and Distance of the Local Group Starburst Galaxy IC 10\altaffilmark{*}}

\author{Minsun Kim\altaffilmark{1,2},
Eunhyeuk Kim\altaffilmark{2,4},
Narae Hwang\altaffilmark{2,3},
Myung Gyoon Lee\altaffilmark{2},
Myungshin Im\altaffilmark{2}, 
Hiroshi Karoji\altaffilmark{3}, 
Junichi Noumaru\altaffilmark{3}, 
and Ichi Tanaka\altaffilmark{3}
}
\email{mskim@kasi.re.kr, ekim@galaxy.yonsei.ac.kr, mglee@astro.snu.ac.kr, \\
narae.hwang@nao.ac.jp, mim@astro.snu.ac.kr, karoji@naoj.org, \\
junichi.noumaru@SubaruTelescope.org, ichi@subaru.naoj.org}

\altaffiltext{1}{International Center for Astrophysics, Korea Astronomy 
and Space Science Institute, Daejeon 305-348,  Korea}
\altaffiltext{2}{Astronomy Program, Department of Physics and Astronomy, 
Seoul National University, Seoul 151-742, Korea}
\altaffiltext{3}{National Astronomical Observatory of Japan,
2-21-1 Osawa, Mitaka, Tokyo, 181-8588, Japan}
\altaffiltext{4}{Institute of Earth-Atmosphere-Astronomy, Yonsei University, Seoul
120-749, Korea}
\altaffiltext{}{*Based on data collected at Subaru Telescope, which is operated by the 
National Astronomical Observatory of Japan.}

\begin{abstract}
We estimate the reddening and distance of the 
nearest starburst galaxy IC 10 using 
deep near infrared $JHK_{S}$ photometry obtained with the 
Multi-Object InfraRed Camera and Spectrograph (MOIRCS)
on the Subaru telescope.
We estimate the foreground reddening toward IC 10 using $UBV$ photometry
of IC 10 from the Local Group Survey, obtaining $E(B-V)=0.52\pm 0.04$ mag.
We derive the total reddening including the internal reddening, 
$E(B-V)=0.98\pm 0.06$ mag, using $UBV$ photometry of early-type stars in IC 10
and comparing $JHK_{S}$ photometry of red giant branch stars in IC 10 and the SMC.
Using the 2MASS point source catalog of 20 Galactic globular clusters, 
we derive a relation between the metallicity [Fe/H]$_{CG97}$ and the slope 
of the red giant branch in the $K_{S}- (J-K_{S})$ color-magnitude diagram. 
The mean metallicity of the red giant branch stars in IC 10 
is estimated to be [Fe/H]$_{CG97}=-1.08\pm0.28$.
The magnitude of the tip of the red giant branch (TRGB) 
of IC 10 in the $K_{S}$ band is measured to be $K_{S,TRGB}=18.28\pm0.01$.  
Based on the TRGB method, 
we estimate the distance modulus of IC 10 to be
$(m-M)_{0}=24.27\pm0.03{\rm (random)}\pm0.18{\rm (systematic)}$,
corresponding to the distance of $d=715\pm10\pm60$ kpc. 
This confirms that IC 10 is a member of the Local Group.

\end{abstract}

\section{Introduction}

The dwarf irregular galaxy IC 10 is the nearest starburst galaxy so that 
it plays an important role for understanding the formation and evolution 
of stars in the starburst galaxies. 
It  is located at the low galactic latitude ($l=118^\circ.95$, $b=-3^\circ.33$) where
the foreground reddening is expected to be large. 
In addition,  the internal reddening of IC 10 may vary spatially due to
its strong star forming activity. 
Therefore, it is difficult to determine reliably the reddening and distance of IC 10,
and published values for reddening and distance of IC 10 
have a large range: $E(B-V)=0.5$ to 2 and $d=500$ kpc to 3 Mpc,
respectively\citep{mas95,sah96,wil96,sak99,bor00,hun01,dem04,vac07,san08,gon08}. 
During the last decade, these rangeg were decreased but are still large:
$E(B-V)=0.8$ to 1.2  and $d=500$ kpc to 1 Mpc  (see \citet{dem04}
and Table \ref{tbl-other} in this study for details).

Since \citet{may35} found that IC 10 is an extra-galactic object,
there were several studies to determine the distance to IC 10. 
In the early days  the distance to IC 10 was estimated to be from 1.3 Mpc to 3 Mpc, 
based on the size of the HII regions
(e.g., \citet{rob62}; \citet{dev65}; \citet{dev78}; \citet{san74}).
\citet{jac81} determined the upper limit of the distance to IC 10 to be 1.8 Mpc
($(m-M)_{0}=26.28$)
using the planetary nebula as a standard candle.
Using the brightest red supergiants and blue supergiants,
\citet{kar93} derived a distance modulus of $(m-M)_{0}=25.08$ ($d=1.04\pm0.09$ Mpc).
\citet{mas95} used the Wolf-Rayet stars and the bluest stars to
obtain a distance modulus of $(m-M)_{0}=24.9$ ($d=0.96$ Mpc). 
\citet{sah96} and \citet{wil96} used the Cepheid variables 
to obtain distance moduli of
$(m-M)_{0}=24.59\pm0.30$ ($d=0.83$ Mpc) and $(m-M)_{0}=24.57\pm0.21$ 
($d=0.82\pm0.08$ Mpc), respectively.
\citet{sak99} used the Cepheid variables
and the $I$ band magnitude of the tip of the red giant branch (TRGB) stars to derive 
a distance of $(m-M)_{0}=24.1\pm0.2$ ($d=600\pm60$ kpc)
and $23.5\pm0.2$ ($d=500\pm50$ kpc), respectively.

\citet{bor00} have determined a reddening of $E(B-V)=1.05\pm0.10$ 
and a distance modulus of $(m-M)_{0}=23.86\pm0.12$ for IC 10
by comparing the $JHK$ phtometry of the red supergiants of IC 10 and IC 1613.
\citet{hun01} used the TRGB to determine a distance to IC 10 of $(m-M)_{0}=24.95\pm0.2$
adopting a reddening of $E(B-V)=0.77$, using $F555W$ and $F814W$ data obtained 
using the Hubble Space Telescope (HST)/Wide Field and Planetary Camera 2 (WFPC2). 
\citet{dem04} determined a distance modulus of $(m-M)_{0}=24.35\pm0.11$
($741\pm37$ kpc), using the mean apparent magnitude of the C stars. 
Using the luminosity function of planetary nebula,
\citet{kni08} obtained a distance modulus of $(m-M) = 24.30^{+0.18}_{-0.10}$.
\citet{vac07} derived a distance modulus of $(m-M)_{0}\sim24.5$
by comparing the location of C stars in the ([F814W]-$K'$)-$K'$ color-magnitude diagram (CMD)
of IC 10 with those of the SMC.
They also derived a distance modulus of $(m-M)_{0}=24.48\pm0.08$
by comparing the location of the TRGB in the CMD of IC 10 with that of the SMC. 
\citet{san08} obtained a distance modulus  of $(m-M)_{0}=24.51\pm0.08$ by comparing the
TRGB magnitude of IC 10 to those of the SMC, 47~Tuc, and $\omega$ Cen, using the deep HST/WFPC2 and
Advance Camera for Surveys (ACS) observations.

The purpose of this paper is to estimate the foreground/total reddening 
toward IC 10 and the distance to IC 10 with high accuracy
using deep near infrared (NIR) $JHK_{S}$ photometry
obtained with the Subaru telescope and also using the optical photometry
of IC 10 in the literature. 
Accurate estimation of the distance to IC 10 is important to understand the nature
of IC 10.
This paper is composed as follows.
\S 2 describes the observation and data reduction, and \S 3 presents the $K_S-(J-K_S)$ CMDs of stars in IC 10.
\S 4 estimates the foreground reddening toward IC 10 and the total reddening of IC 10.
In \S 5, we derive a relation between metallicity [Fe/H]$_{CG97}$ and the red giant branch (RGB) slope in the $K_S-(J-K_S)$ CMD,
and we use it to estimate the mean metallicity of the RGB stars in IC 10.
Then we determine the distance to IC 10 using the $J,H,$ and $K_S$ band magnitudes of the TRGB. 
In \S 6, we compare our results to those in the previous studies.
Primary results of this study are summarized in \S 7. 
Throughout this paper, quoted errors are for $\pm1\sigma$ confidence level. 

\section{Observation and Data Reduction}

\subsection{Observation}

Deep imaging observations were carried out on 2005 December 9,  with 
Multi-Object InfraRed Camera and Spectrograph (MOIRCS)
\footnote{http://www.subarutelescope.org/Observing/Instruments/MOIRCS/index.html}, 
a wide-field imaging camera and spectrograph on
the Subaru telescope. 
MOIRCS consists of two HAWAII-2 HgCdTe arrays 
and has a field of view of 
$4\arcmin\times7\arcmin$ with a spatial 
resolution of $0.117\arcsec/\rm{pixel}$.

We observed a field centered at IC 10 (R.A. [J2000.0]$=00^{h}20^{m}17.303^{s}$, 
Decl. [J2000.0]$=+59^{o}18{\arcmin}13.66{\arcsec}$) with a dithering mode 
using the $J$, $H$, and $K_{S}$ band filters
\footnote{http://www.subarutelescope.org/Observing/Instruments/MOIRCS/imag\_sensitivity.html}.
Our observation log is given in Table 1. 
Total exposure times are 900 s, 675 s, and 1215 s for the
$J$, $H$, and $K_{S}$ bands, respectively. 
Seeing was excellent during the observations, 0\arcsec.5 to 0\arcsec.6.

\subsection{Data Reduction}

Pre-processing of the images was performed in the standard manner by subtracting dark and bias 
and by performing flat correction using IRAF/XDIMSUM.
We corrected the optical distortion of the images 
that is fit well by the third-order polynomial as a function of the distance 
from the optical center ($x_c=858$, $y_c=1034$ on chip 1, $x_c=1178$, $y_c=1012$ on chip 2)
\footnote{http://www.subarutelescope.org/Observing/Instruments/MOIRCS/imag\_information.html},
where $x_c$ and $y_c$ are coordinates of the center in each chip.
Individual long exposure images were combined with 
XDIMSUM\footnote{http://iraf.noao.edu/iraf/ftp/iraf/extern-v212/xdimsum}
to make a single mosaic image and to remove the varying sky fluctuation in each band. 

Figure \ref{fig-ic10region_dpi150}
shows a gray scale map of the $K_{S}$ band image of IC 10. 
We note that numerous resolved stars are visible throughout the field 
with a higher concentration of stars around the middle part of the field,
and that little signature of dust feature is visible.

We divided the entire observed field (called RA)  into 
three subregions for the following analysis as shown in Figure \ref{fig-ic10region_dpi150}: 
R2 for the central region, and R1 and R3 for the northern and southern outer regions,
respectively.
We listed the mean surface number density of the stars for each region in Table \ref{tbl-region}.
The mean surface number density of stars in the R2 region is $\sim0.7/$arcsec$^{2}$, 
and that of the R1 and R3 regions is  $\sim0.4/$arcsec$^{2}$.
Therefore, there are a factor of 1.8 times more stars per unit area in the central region compared to
the outer regions.  

Instrumental magnitudes of point sources in the images were derived from the combined images for each filter, using  a custom software
{\bf $alfphot$}, which runs automatically  
a combination of DAOPHOT, ALLSTAR, MONTAGE2, and ALLFRAME \citep{ste94}.
We transformed the instrumental magnitudes onto standard magnitudes 
using $JHK_{S}$ photometry in the Two Micron  
All Sky Survey (2MASS)
\footnote{http://www.ipac.caltech.edu/2mass} point source catalog
for the objects that are common with those in our observation.

We selected several dozens of bright stars in the 2MASS point source catalog with
good quality flags (AAA), and matched these sources with those in our observation.
Due to the large difference in seeing FWHMs between 2MASS images($\sim 3\arcsec$) and the present
SUBARU $JHK_S$ images ($\sim 0.5\arcsec$) 
we consider the matches with matching distance smaller than $0.9\arcsec$.
Figure \ref{fig-calib} displays the difference between the 2MASS magnitudes and the instrumental magnitudes
for these stars.
We used the data points inside the boxes in each panel for calibration.
The mean values for the differences (the 2MASS magnitudes minus the instrumental magnitudes in this study)
are: $2.92 \pm 0.06$ mag (31 stars),   $2.99 \pm 0.09$ mag (29 stars), $2.20 \pm 0.07$ mag (37 stars),
for $J$, $H$, and $K_S$, respectively.
Therefore the mean errors for calibration are  0.010, 0.017, and 0.012
 for $J$, $H$, and $K_S$, respectively.
It is known that the effect of color term is negligible between 2MASS photometric system 
and the NIR photometric system in Mauna Kea Observatories \citep{leg06} 
so that we did not consider any color term for standard calibration.

The final catalog of $JHK_S$ photometry of IC 10 
includes $\sim52,000$ stars.
The magnitude limits are $J\sim22$, $H\sim21.5$,
and $K_{S} \sim21$ mag, respectively. 
In Figure \ref{fig-mag_err}, we display mean
photometric errors of the stars in the $K_{S}$, $(J-K_{S})$, and $(J-H)$  
as a function of the $K_{S}$ magnitude.  
The photometric errors in Figure 3 represent the mean of the photometric
errors of DAOPHOT/ALLFRAME photometry at a given magnitude range.
These errors were derived from the photometry of the combined images for each filter hence 
represent the photon noise. We also checked the errors due to frame-to-frame repeatability using the photometry of individual images, finding that
they are similar to those derived from the combined images.

\subsection{Comparison with Previous Photometry}

\citet{bor00} presented $JHK$ photometry of bright stars 
in a central $3\arcmin.6 \times 3\arcmin.6$ field of IC 10.
We compared our photometry with that of \citet{bor00},
displaying the magnitude differences between this study and \citet{bor00} 
in Figure \ref{fig-comp_bori}.
The mean magnitude differences (this study minus \citet{bor00}) with $2\sigma$ clipping 
are $+0.23$, $+0.29$, and $+0.11$ mag for the $J$, $H$, and $K_{S}$ bands, respectively. 
If we use only the bright stars with  $J<17$, $H<16$, and $K_S<16$ mag, 
these differences decrease  significantly:
$+0.14$, $+0.18$, and $+0.05$ mag for the $J$, $H$, and $K_{S}$ bands, respectively. 
Our magnitudes are slightly fainter than \citet{bor00}'s.

To understand the cause of this magnitude difference,
we estimate the effect of source blending due to the difference in the spatial resolution. 
Note that the data used for \citet{bor00} were obtained
under the seeing of $1\arcsec-1\arcsec.2$ and the scale of the NIR detector
of $0\arcsec.85$/pixel. 
To estimate the degree of brightening of a source due to the unresolved nearby faint point
sources, we first compared the luminosity function of point sources in the region common between
the two studies.
For the bright point sources with $K_{S}<16$ mag, there are 219 point sources in \citet{bor00},
which is similar to the number of sources found in this study, 295.
For the fainter sources with $16<K_{S}<17$ mag,
we found 295 point sources in \citet{bor00}, which is much smaller than the number of sources found in this study, 739.
This is consistent with the incompleteness results given in \citet{bor00}. 

We assume that the faint point sources with $K_{S}>17$ mag,  which were
completely or partially detected in this study, are merged into the nearby brighter stars with $K_{S}<16$ mag
due to the poor spatial resolution. 
 With this assumption and the seeing FWHM of \citet{bor00},  
we computed the mean brightness of those sources and found that the mean brightness of
brighter stars ($J<17, H<16$ mag, and $K_{S}<16$ mag) increases by $0.106(J), 0.057(H),$ and $0.049(K_{S})$.
With these correction values the magnitude differences between ours and \citet{bor00} decrease
to small values of $0.03(J), 0.12(H)$, and $0.00(K_{S})$, respectively.
Therefore the magnitude differences between the two studies are mainly due to difference in the spatial
resolution of the images.
We note that the $JHK_{S}$ photometry of this study is significantly deeper ($\sim$ 2 mag) 
than that of \citet{bor00}. Therefore a large scatter at fainter magnitudes ($K_{S}>16$ mag)  in Figure 4
is mainly due to the large photometric errors in \citet{bor00}.

\section{Color-Magnitude Diagrams}

In Figure \ref{fig-cmd_jkk_con_n3}, 
we display the $K_{S} - (J-K_{S})$ CMD of the measured stars in  
 the entire observed region  and three subregions. 
The number density contour maps are overplotted to show clearly
 the morphology of the RGB.
Several features are noted in Figure \ref{fig-cmd_jkk_con_n3}.
First, the most distinguishable feature is an RGB population with $1<(J-K_{S} ) <2$, 
showing that most of the resolved stars are red giants with $K_{S} = 18.3 - 20.5$ mag.
Second, there is an additional notable feature brighter  than the 
brightest part of the RGB, which is an AGB population. 
AGB stars have $K_S$ magnitudes of 17 to 18 mag.
Third, there is a horizontal feature extended to the redder color 
from the brightest part of the AGB, which are mostly carbon stars.
Fourth, there is an almost vertical bright sequence with $0.5\lesssim(J-K_{S})\lesssim1$,
which are mostly foreground main-sequence (MS) stars belonging to the Galaxy.

\section{Reddening of IC 10}

We estimate both foreground and internal reddening of IC 10 
using $UBV$ photometry from the Local Group Survey given by \citet{mas07} and
$JHK_S$ photometry in this study. 
We adopted the extinction laws for $R_V=3.3$, 
$A_{J}/A_V=0.2957$, 
$A_{H}/A_V=0.1869$, and 
$A_{K_S}/A_V=0.1206$ 
\citep{car89}. 

\subsection{Foreground Reddening}

We estimate the foreground reddening toward IC 10 using  $UBV$ photometry of 
the foreground stars in the direction of IC 10.
We use $UBV$ photometry of a $20\arcmin \times 30\arcmin$ field covering IC 10
given in the Local Group Survey \citep{mas07}. 
Figure \ref{fig-plot_ic10_ext4paper1}(a) displays the optical $V-(B-V)$ CMD of these stars. 
The majority of point sources seen in the CMD are foreground stars. 
The blue plume fainter than $V\approx 18$ mag in the CMD represents mainly the bright MS stars in IC 10.
To estimate the foreground reddening toward IC 10, 
we selected bright  foreground MS stars  
inside the large tilted box in Figure \ref{fig-plot_ic10_ext4paper1}(a). 

 Figure \ref{fig-plot_ic10_ext4paper1}(b) shows the $(U-B)-(B-V)$ diagram of these selected foreground MS
stars.
We adopted $E(U-B)/E(B-V)=0.72$, and showed the reddening direction for one magnitude
visual extinction by an arrow in Figure \ref{fig-plot_ic10_ext4paper1}(b).
Recently \citet{zag07} argued that there might be insignificant difference in
the extinction law between the Milky Way, LMC, and SMC.
We binned the data in $(B-V)$ color with a magnitude bin size of $0.05$  
and fitted these with the empirical fiducial relation 
for the MS  \citep{sch82} using the
chi-square minimization method. 
We obtained the best fit 
value  of $E(B-V)=0.52\pm0.04$ with $90\%$ confidence level. 
This value  is $\sim 3$ times smaller  than that in \citet{sch98}, $E(B-V)=1.527$.
It is noted that the reddening values for the low galactic latitude in \citet{sch98}
have a large uncertainty.

Due to the large sky area of the Local group Survey \citep{mas07} 
and the low galactic latitude of IC 10, 
there might be spatial variation of the foreground reddening. 
To investigate any variation of the foreground reddening, we divided the sample
of the foreground MS stars in the Local Group Survey 
into four groups with different galactocentric distance ($r$).
Using the same method applied to estimate the foreground reddening for the total sample,
we obtained the values:
$E(B-V) = 0.56 \pm 0.08$, $0.52 \pm 0.06$, $0.50 \pm 0.05$, and $0.52 \pm 0.07$ for $r<5\arcmin$, 
$5\arcmin<r<8\arcmin$, $8\arcmin <r<10\arcmin$ , and $10\arcmin<r<13\arcmin$, respectively. 
This result shows that there is little spatial variation of the foreground reddening over the field in this study.
Therefore we adopt the foreground reddening derived using the total sample of \citet{mas07} as
a foreground reddening toward IC 10.

The value for the foreground reddening of IC 10 derived above would be a lower limit for the following reasons.
First, the $(U-B)$ color distribution of the foreground MS stars used for the reddening estimate shows a large dispersion 
of $\sigma=0.13$, which is $\sim4$ times larger than the mean photometric error.
Second, the foreground reddening was estimated by using the
mean colors of MS stars rather than the colors of individual stars. 
This large dispersion in the color of stars could be generally caused by 
1) different galactic stellar population with different metallicity toward IC 10,
2) photometric errors of observation, or
3) differential reddening of individual stars in the direction of IC 10. 
Because we observed a field in the very low galactic latitude ($b=-3^\circ.33 $), 
we may see through the Galactic thin disk
in which the stellar population may be somewhat homogeneous, having similar metallicity and age.
Therefore, the effect of the stellar population for the large color dispersion may be negligible.
The photometric uncertainty should not be a cause for the large color dispersion in IC 10,
because we considered only the bright stars with small photometric errors.  
Therefore, we conclude that the large dispersion in 
the $(U-B)$ color of the foreground stars is mainly attributed  
to the differential reddening of stars toward IC 10.

Using the relation between $UBV$ photometric system and MK spectral
classification of \citet{sch82} 
and the derived foreground reddening, $E(B-V)=0.52$,
we computed the $V$ band distance moduli of the foreground
MS stars in the tilted box of Figure 6(a). 
We found that ~90\% of foreground stars
are within $\sim7$ kpc from the Sun with a median distance of $\sim3.7$ kpc,
consistent with the galactic longitude of IC 10, $l = 118^\circ.95$. 
We also found that the faintest ($V\approx22$ mag) red MS stars are located at the similar distance
to that of the brightest ($V\approx16$ mag) blue MS stars, 
while there is a significant spatial spread for stars with intermediate colors.

\subsection{Total Reddening}

The total reddening of IC 10 consists of the foreground reddening (due to the interstellar dust 
in the Milky Way) and the internal reddening (due to the interstellar dust in IC 10). 
We applied two different approaches to estimate the total reddening:
1) comparing the mean NIR colors of RGB stars in IC 10 
and those in the SMC in the $K_S-(J-K_S)$ CMD,
and 2) using the $(U-B)-(B-V)$ diagram of the early-type stars in IC 10.

\subsubsection{NIR Colors of the RGB stars}

Since the mean NIR color of the RGB stars in a galaxy is expected to be similar to that of another 
galaxy with a similar metallicity,
we can use the mean NIR color of the RGB stars to estimate the reddening of a galaxy.
We selected the SMC as a comparison with IC 10, because the SMC, a nearby dwarf irregular galaxy, 
is known to have a similar metallicity to that of IC 10 \citep{cio00a}.

We used $JHK_S$ photometry of the SMC stars given by \citet{kat07}.
Their photometry is given in the 2MASS system, and 
reaches $\sim3$ mag deeper than the 2MASS.
We compared the mean NIR colors of the RGB stars of IC 10 and the SMC as follows, 
assuming that the difference in metallicity between two galaxies is small ($\lesssim0.2$ dex).
Because our $JHK_S$ photometry of IC 10 includes stars $\sim2.0$ mag 
fainter than the TRGB, we compared the mean colors of RGB stars 
with $K_{S,TRGB}<{K_S}<K_{S,TRGB}+1.5$, 
where $K_{S,TRGB}$ is an apparent magnitude of the TRGB in the $K_{S}$ band. 

The TRGB magnitude for IC 10 is $K_{S,TRGB}=18.28$ mag 
 as derived in \S 5. 
We estimated the $K_{S,TRGB}$ of the SMC as follows.
To take advantage of multi-band photometry, we select the candidate
RGB and AGB stars with negligible internal reddening in the SMC 
using both optical and NIR photometry of the SMC.
Using the deep optical photometry of the SMC given by \citet{zar02}, 
we selected RGB and AGB stars in the outskirt of the SMC 
where the internal reddening for the RGB stars is expected to be negligible \citep{zar02}. 
We matched the selected RGB and AGB stars 
to those in the NIR catalog of \citet{kat07},
obtaining an NIR photometric list of RGB and AGB stars in
the outer region of the SMC.
With these RGB and AGB candidates, we determine the magnitude of the TRGB in the SMC 
to be $K_{S,TRGB}=12.85$ mag. 
This value is very similar to that given in \citet{kat07}, $K_{S,TRGB}=12.80$ mag.
However, this value is  about 0.2 mag fainter than that given by \citet{cio00a}, $K_{S,TRGB}=12.62$ mag, 
who used the DENIS catalog towards the Magellanic Clouds \citep{cio00b}. 

We derived the mean loci of RGB stars in IC 10 and SMC with $0.2$ magnitude 
interval in the $K_S$ band, 
and estimated the mean and dispersion of the $(J-K_{S})$ colors by fitting the color  distributions of RGB stars with a Gaussian function. 
Figure \ref{fig-plot_ic10_ext4paper2} displays the $K_{S} - (J-K_{S})$
CMD of IC 10 with the mean locus of RGB stars for IC 10 ({\it{solid line}})
and that for the SMC ({\it{filled circles}}). 
The RGB locus for the SMC was shifted vertically so that the TRGB of the SMC matches that of IC 10.
Then the RGB locus of the SMC was shifted horizontally so that it matches that of IC 10.
We found an excellent agreement between two RGB loci.
The difference in reddening between IC 10 and the SMC thus derived is 
$\Delta E(J-K_{S})$(IC 10--SMC)$=0.56\pm0.02$ mag, and 
$\Delta E(B-V)$(IC 10--SMC)$=0.97\pm0.03$ mag. 
Here the error  represents the mean error of the derived mean color.

It is noted that the RGB slope of IC 10 is very similar to that of the SMC,
supporting the assumption employed in this analysis.
Adopting the foreground reddening of the SMC of $E(B-V)=0.04$ given in \citet{sch98},
we derive a value for the total reddening of the RGB stars in IC 10, $E(B-V)=1.01 \pm 0.03$.

\subsubsection{UBV photometry of Bright Blue Stars of IC 10}

Using the $UBV$ photometry of bright blue MS stars in IC 10 from the 
Local Group Survey \citep{mas07}, 
we directly estimated the total reddening for IC 10. 
We selected bright blue stars in IC 10 with the conditions:
$B<22.5$ mag, $-1.0<(U-B)<0.2$, and $0.3<(B-V)<1.0$.
These stars were plotted as triangles in Figure 6. 
We obtained the reddening of individual
stars comparing the empirical fiducial line of the MS \citep{sch82} and the selected bright blue stars
in the $(U-B)-(B-V)$ diagram, as shown in Figure 6(b).
The mean value for the total reddening is measured to be $E(B-V)=0.95\pm0.06$. 
This value is in good agreement 
with the total reddening of $E(B-V)=1.01 \pm 0.03$  derived using NIR colors of the RGB stars.

Taking an average of the two estimates, 
we derive a value of $E(B-V)=0.98 \pm 0.06$ 
for the total reddening for IC 10, which is used in the subsequent analysis. 
This result is similar to 
the values, $E(B-V)=1.05\pm0.10$ given by \citet{bor00} 
and $E(B-V)=1.16$ given by \citet{sak99}, but
it is larger than other previous estimations based on various observations and methods (see Table \ref{tbl-other}.).
It is  noted that \citet{mas07} derived a reddening value of $E(B-V)=0.81$ for IC 10,
which is  smaller than our estimation, 
although both studies used the same photometric catalog.

\section{Distance to IC 10}

\subsection{The RGB Slope and Metallicity}

\citet{lee93} suggested that the absolute magnitude of the TRGB in the $I$-band
is a primary distance indicator for resolved galaxies since it depends little on
metallicity ([Fe/H]$<-0.7$ dex) and age ($>$a few Gyrs)
of old stellar populations.
However, the absolute magnitudes of the TRGB  depend 
on metallicity 
much more in the $JHK_{S}$ bands than in the $I$ band  \citep[e.g.][]{val04,fer06}.
Therefore we need to know the metallicity of RGB stars when using the NIR TRGB magnitudes
for the distance estimation.

It is known that the RGB slope of old stellar populations in the CMD is
sensitive to the metallicity, while insensitive to age, 
and that the RGB slope has a strong correlation with metallicity \citep[e.g.][]{kuc95a,kuc95b,mig98}.
Therefore the RGB slope can be used to estimate metallicity of old stellar populations.
Since the slope of the RGB does not depend either on reddening or on distance,
the mean metallicity of the RGB stars estimated from the RGB slope
is independent of the reddening and distance of RGB stars.
The RGB slope can be measured 
using RGB stars in a full range of magnitudes or in a specific magnitude range: 
RGB stars with brightness of the top of 
horizontal branch (HB) to 4.6 mags brighter in the $K-(J-K)$ CMD ((\citet{kuc95a}; \citet{kuc95b}) or
stars on the RGB from top of the HB to 2.0 and 2.5 mags brighter 
in the $V-(B-V)$ CMD \citep{mig98}.

We derived a relation between the RGB slope in the $K_{S} - (J-K_{S})$ CMD
and the metallicity of Galactic globular clusters, where the RGB slope is measured using the
RGB stars in the bright $\sim 2$ magnitude range as well as in several other magnitude ranges.
We selected a photometric sample of 20 Galactic globular clusters as calibrators from
the 2MASS point source catalog. 
The members of each globular cluster were selected inside 2 to 4 times of the half mass radii ($r_{h}$)
of the globular cluster depending on the tidal radii ($r_t$) of clusters,
where the outer boundary is chosen to minimize the background
contamination and to maximize the number of the cluster members.
We note that the basic parameters of Galactic globular clusters such as positions and $r_{h}$
are based on the literature \citep{har96}, unless otherwise noted.
In addition, we selected stars having 2MASS $rd\_flg$ values of 1, 2, or 3,
which indicate the best quality detections, photometry, and astrometry.
With these selection criteria, the number of stars in each globular cluster is $100\sim3000$
depending on the apparent size of a globular cluster.
We derived the mean locus of the RGB stars in the $K_S - (J-K_{S})$ CMD 
by iterative fitting with second order polynomial equations clipping the stars deviating by $3\sigma$ from the fitted line. 
We determined the position of the TRGB by careful visual inspection.
We estimated the RGB slopes in several magnitude steps,
between the magnitude of the TRGB and the magnitudes fainter than that of the TRGB,
using the mean locus of the globular cluster.

In Table \ref{tbl-gc} we list the parameters of the Galactic globular clusters
that were used for calibrating the relation between the RGB slope and the metallicity.
We note that the values of [Fe/H]$_{CG97}$ \citep{car97} come from the literature
such as \citet{har96}, \citet{fer99,fer00}, and \citet{val07}.
We transformed  [Fe/H]$_{ZW}$ into [Fe/H]$_{CG97}$ using the equation (7)
in \citet{car97}, [Fe/H]$_{CG97} = -0.618 -0.097$[Fe/H]$_{ZW} - 0.352$[Fe/H]$^2_{ZW}$.

The RGB slopes as a function of [Fe/H]$_{CG97}$ \citep{car97}
are displayed for several magnitude steps  in Figure \ref{fig-slopefeohfit}.
The best linear least-squares fit results of the relation between
${\rm [Fe/H]}_{CG97}$ and the RGB slope are derived as follows:

\begin{equation}
{\rm [Fe/H]}_{CG97,S_{05}} = 0.09(\pm0.02)S_{05}-0.48(\pm0.19)~ [\sigma=\pm0.34]
\end{equation}
\begin{equation}
{\rm [Fe/H]}_{CG97,S_{10}} = 0.10(\pm0.02)S_{10}-0.36(\pm0.18)~ [\sigma=\pm0.31]
\end{equation}
\begin{equation}
{\rm [Fe/H]}_{CG97,S_{15}} = 0.11(\pm0.02)S_{15}-0.25(\pm0.17)~ [\sigma=\pm0.28]
\end{equation}
\begin{equation}
{\rm [Fe/H]}_{CG97,S_{20}} = 0.11(\pm0.02)S_{20}-0.16(\pm0.16)~ [\sigma=\pm0.26]
\end{equation}
\begin{equation}
{\rm [Fe/H]}_{CG97,S_{25}} = 0.11(\pm0.01)S_{25}-0.09(\pm0.15)~ [\sigma=\pm0.23]
\end{equation}
\begin{equation}
{\rm [Fe/H]}_{CG97,S_{30}} = 0.11(\pm0.01)S_{30}-0.02(\pm0.15)~ [\sigma=\pm0.22]
\end{equation}
\begin{equation}
{\rm [Fe/H]}_{CG97,S_{35}} = 0.11(\pm0.01)S_{35}+0.03(\pm0.17)~ [\sigma=\pm0.20]
\end{equation}
\begin{equation}
{\rm [Fe/H]}_{CG97,S_{40}} = 0.11(\pm0.01)S_{40}+0.07(\pm0.13)~ [\sigma=\pm0.19]
\end{equation}

\noindent where $S_{05}$, $S_{10}$, $S_{15}$, $S_{20}$, $S_{25}$, $S_{30}$,
$S_{35}$, and $S_{40}$ are linearly fitted slopes in magnitude ranges
between the TRGB
and 0.5, 1.0, 1.5, 2.0, 2.5, 3.0, 3.5, and 4.0 magnitude fainter than
the TRGB, respectively.
The uncertainties for parameters and fitting errors are given in parentheses and brackets,
respectively.
These uncertainties decrease with increasing the magnitude ranges of RGB stars.

Our photometric catalog of IC 10 reaches only $\approx 2$ magnitude 
fainter than the TRGB.
Therefore, we estimated the mean metallicity of the bright RGB stars in IC 10 
by computing $S_{15}$ and using equation (3) (see Table 4).
The mean metallicity of the RGB stars in the entire region (RA) is
measured to be [Fe/H]$_{CG97}$ $=-1.08\pm0.28$.
The [Fe/H]$_{CG97}$ of the central region (R2) of IC 10 ([Fe/H]$_{CG97} =-1.06\pm0.28$)
is very similar with that of outer regions (R1 and R3) ([Fe/H]$_{CG97} =-1.05\pm0.28$).
The metallicity of the RGB stars corresponding to intermediate-age and old populations in IC 10
is expected to be smaller compared to that of the young stellar population such as HII regions.
\citet{gar90} presented a value for the [O/H], 12+log[O/H]=8.19, derived from the analysis of the data for the HII regions in IC 10. 
This corresponds to a metallicity of [Fe/H]$_{CG97} =-0.8$ \citep{bat05}, 
which is higher than that of RGB stars estimated in this study.
We note that the young stellar population hardly affects the determination of
the RGB slope and TRGB for IC 10, because the number of young stars in the RGB region in the CMD is negligible.
We matched the young stellar populations (triangles in the Figure \ref{fig-plot_ic10_ext4paper1})
with $JHK_{S}$ photometry, and found only $\sim350$
matched sources. 
This  number is negligible compared with the number of the RGB stars used in estimating the RGB slopes and the TRGB.

\citet{cio00a} determined [M/H] of the LMC and SMC by fitting isochrones \citep{gir00} to
the $K_S - (J-K_{S})$ CMD,
finding [M/H]=$-0.70$ and [M/H]=$-0.82$ for the LMC and SMC, respectively.
We transform [M/H] to [Fe/H]$_{CG97}$ using an equation of
[Fe/H]$_{CG97}$=[M/H]$-log(0.638\times10^{0.28}+0.362) = [M/H] - 0.20$\citep{fer99},
resulting in [Fe/H]$_{CG97}=-1.02$ for the SMC.
Therefore, the mean metallicity of RGB stars in IC 10 determined in this study
is similar to that of the SMC estimated by \citet{cio00a}.
In addition, we estimate the mean metallicity of the SMC using the RGB slope in the
$K_S - (J-K_{S})$ CMD obtained from \citet{kat07} (see Figure \ref{fig-plot_ic10_ext4paper2}).
Adopting the equation (3) for $S_{15}$ and (6) for $S_{30}$, 
the mean metallicity of the RGB stars in the SMC is derived to be
[Fe/H]$_{CG97}=-1.15\pm0.28$ and [Fe/H]$_{CG97}=-1.03\pm0.22$, respectively.
These values agree well with that of the SMC given by \citet{cio00a}.
Thus the metallicity of the RGB stars in IC 10 is very similar to that of the SMC.

\subsection{Distance}

We estimate the distance to IC 10 using the TRGB method \citep{lee93}.
First we measure the apparent magnitude of
the TRGB ($m_{TRGB}$) at which the luminosity function of red giant stars shows a sudden increment.
Figures 9, 10 and 11 display the luminosity functions $N(m)$ of the red giant stars in IC 10
for $J$, $H$, and $K_S$ bands, respectively.
These figures show two types of increments: a slow increment due to the AGB stars 
in brighter magnitudes, and a rapid increment due to the RGB stars in fainter magnitudes.
The latter represents the TRGB. 
The TRGB is seen at  $J\approx 19.8$ mag,  $H\approx 18.7$ mag, and ${K_S}\approx 18.3$ mag, respectively.

To quantitatively determine the magnitude of the TRGB, 
we derive the second derivative of the luminosity function, $N''(m) \equiv d^{2}N(m)/dm^{2}$,
following the method suggested by \citet{cio00a}, as follows:
1) we derive the luminosity function $N(m)$ for the red stars,
2) we apply a Savitzky-Golay filter \citep{pre92}    
to derive $N''(m)$,
3) we find the highest peak of $N''(m)$,  fitting it with a Gaussian function,
and 
finally we estimate an apparent magnitude of the TRGB,
$m_{TRGB}=m_{2g}-\Delta m_{2g}(\sigma_{2g})$ 
where $m_{2g}$ and $\sigma_{2g}$ are the mean and dispersion of the best fit Gaussian.
$\Delta m_{2g}(\sigma_{2g})$ is a correction factor as a monotonic function of $\sigma_{2g}$
due to a  phenomenological model. 
We used the solid line ($\Delta f=0.25$) in Figure A.2 (b) of \citet{cio00a}
to estimate $\Delta m_{2g}(\sigma_{2g})$, 
assuming that the shape of the intrinsic magnitude distribution of the stars in IC 10 is not significantly
different from that of the LMC or SMC.
The estimated values for $\Delta m_{2g}(\sigma_{2g})$ for the $JHK_{S}$ bands in all regions have a range 
from $-0.09$ to $-0.14$.  Details of this method are given in Appendix of \citet{cio00a}. 

We performed Monte-Carlo simulations to estimate the uncertainties of
the derived value for $m_{TRGB}$. We generated a thousand of random realizations having 
the same number of stars as the observation. 
The magnitude of an artificial star in the random realization was randomly drawn from a 
Gaussian distribution with a Gaussian width of the photometric uncertainty centered
at an observed stellar magnitude.
We perform the same processes applied for the observed data 
to every random realization to detect the TRGB magnitude,
and then calculate the median, $\tilde m_{TRGB,sim}$, and 
dispersion, $\sigma_{TRGB,sim}$, of the TRGB magnitudes. 
The resulting $\tilde m_{TRGB,sim}$ shows an excellent agreement with the observation value $m_{TRGB}$ 
($<0.01$ mag) and the derived $\sigma_{TRGB,sim}$ is smaller than 0.02 magnitude.
We attribute $\sigma_{TRGB,sim}$ as the uncertainty of the derived
 value $m_{TRGB}$. 
The TRGB magnitudes estimated in this study are listed in Table \ref{tbl-dm} for the $JHK_{S}$ bands
and are displayed in Figures 9, 10, and 11, 
which  show $N(m)$ and $N''(m)$ of stellar populations of IC 10 for the $JHK_{S}$ bands,
respectively. 
These values for The TRGB magnitudes are in excellent agreement with the visual estimates
with differences smaller than 0.05.

 \citet{val04} provided empirical calibrations of the absolute
magnitude of the TRGB as a function of [Fe/H]$_{CG97}$, based on a homogeneous
NIR data of 24 Galactic globular clusters with a wide
metallicity range ($-2.16 \leq $[Fe/H]$_{CG97} \leq -0.38$). 
To estimate the distance modulus of IC 10, we compute
the absolute magnitude of the TRGB for each region 
using the following equations 
given by \citet{val04}:
\begin{equation}
M_J=-0.31 {\rm [Fe/H]}_{CG97}-5.67~~[\sigma=\pm0.20]
\end{equation}
\begin{equation}
M_H=-0.47 {\rm [Fe/H]}_{CG97}-6.71~~[\sigma=\pm0.16]
\end{equation}
\begin{equation}
M_{K_{S}}=-0.58 {\rm [Fe/H]}_{CG97}-6.98~~[\sigma=\pm0.18]
\end{equation}
where $\sigma$ is the dispersion of each relation. 
We consider this dispersion as the
uncertainty of the absolute magnitude of the TRGB in this study. 
We note that this uncertainty is much larger than that of  
the apparent magnitude of TRGB ($\lesssim0.02$ mag).
The absolute magnitudes of the TRGB in the entire observed region (RA) 
are $M_J=-5.33\pm0.20$ mag,
$M_H=-6.20\pm0.16$ mag, and $M_{K_S} = -6.35\pm0.18$ mag, respectively. 
Using the absolute magnitude and the apparent magnitude of TRGB stars,
we estimate the distance modulus for IC 10.
In Table \ref{tbl-dm} we list the absolute magnitudes of the TRGB and 
distance moduli derived for the entire and three subregions for the $JHK_{S}$ bands. 
The extinction values used for IC 10 are 
$A_J = 0.96\pm0.03$, $A_H =  0.60\pm0.02$, and $A_{K_{S}}=0.39\pm0.02$ mag.
 
The distance moduli for IC 10 derived for three subregions and three different bands
agree well within the uncertainties.
Since the crowding in outer regions is lower than that in the central region 
and the uncertainty of reddening 
is the smallest in the $K_{S}$ band, we derive a distance modulus of IC 10
by averaging the distance moduli of R1 and R3 regions in the $K_{S}$ band:
$(m-M)_{0}=24.27\pm0.03(random)\pm0.18(systematic)$.
This corresponds to a distance of 
$d=715^{+10}_{-10}$ $^{+62}_{-57}$ kpc. 
We  note that the quoted error of the distance modulus consists of random and 
systematic uncertainties.
The random error comes from the uncertainties of TRGB detection and applied reddening,
while the systematic error comes from the uncertainty in 
calibration equation of the absolute magnitude of TRGB. 
To reduce the uncertainty in the
calibration equation in deriving the absolute magnitude of TRGB stars, 
higher precision $JHK_S$ photometry of the stars in Galactic globular clusters is needed.

\section{Discussion}

The distance to IC 10 has been derived with various standard candles:
the Wolf-Rayet stars and blue plume \citep{mas95}, Cepheid variables 
\citep{wil96, sah96, sak99}), red supergiant stars \citep{bor00},
carbon stars \citep{dem04}, and the TRGB stars 
\citep{sak99,hun01,vac07,san08}.
Previous estimates for the distance modulus of IC 10 range
from $(m-M)_{0}=23.5$ to $25.0$ (see Table \ref{tbl-other}), 
and the value derived in this study, 
$(m-M)_{0}=24.25\pm0.03\pm0.18$, 
is in the middle of the previous estimates.

Since different standard candles suffer from different reddening, it is not simple to figure out
what causes the difference in the estimated distances.
Therefore we focus on the distance estimates only based on the TRGB
 to investigate what causes the difference in the estimated distances. 
It is noted that the distance estimates 
based on the TRGB method also  
show a large dispersion:
$(m-M)_{0}=23.51\pm0.19$ in \citet{sak99}, 
$24.95\pm0.20$ in \citet{hun01},   
$24.48\pm0.08\pm0.16$ in \citet{vac07}, and
$24.51\pm0.08\pm0.08$ in \citet{san08}.

All previous studies based on the TRGB method assumed $R_{V}=3.1$
except for \citet{sak99} who adopted $R_{V}=3.2$, while we adopted  $R_{V}=3.3$.
We checked how much the difference in the assumed value of $R_{V}=A_{V}/E(B-V)$ 
contributes to the distance modulus estimation.
When $E(B-V)\sim1$ and $3.0\leq R_{V} \leq 3.3$ were assumed, 
the various assumption of $R_{V}$ results in the distance scatters of 0.08 and 0.2 mag
at maximum for  NIR ($JHK$) bands and optical ($BVI$) bands, 
respectively.
Therefore, the different value of $R_{V}$ does not contribute much to
the large scatter in distance moduli ($\sim 1$ mag) as seen in Table 5.

\citet{sak99} determined the distance to IC 10 based on the TRGB method
using $VI$ photometry of IC 10 obtained at the Hale 5 m telescope.
They derived $I_{TRGB}=21.70\pm0.15$,
 and obtained a distance modulus of $(m-M)_{0}=23.5\pm0.2$ ($500\pm50$ kpc),
adopting a reddening value of $E(B-V)=1.16\pm0.08$. 
If the reddening value of $E(B-V)=0.98$ is adopted as in this study, 
their distance modulus becomes larger by $0.27$ mag, 
but it is still smaller than ours by $0.50$ mag.
It is noted that their TRGB magnitude is 0.5 mag brighter than that \citet{hun01} 
derived from the $HST$ data. 
Therefore the TRGB in their estimate may be the tip of the AGB,
or their estimate for the TRGB magnitude may be an overestimate due to the blending effect
in the CCD images they used.

\citet{hun01} determined the distance to IC 10 based on the TRGB method
using $F555W$ and $F814W$ photometry obtained from $HST$ observation.
They derived $I_{TRGB}=22.2$, and estimated a distance modulus of $(m-M)_{0}=24.9$ (0.95 Mpc)  adopting a reddening of $E(B-V)=0.77$. 
Their distance modulus is $0.63$ magnitude larger than that of this study. 
If the same reddening of $E(B-V)=0.98$ as in this study is used,
their distance modulus will be only 0.09 magnitude larger than ours,
agreeing well with our estimate. 

\citet{vac07} obtained both optical and NIR photometry 
of IC 10 from the laser guide star adaptive optics (AO)  observation 
at the Keck II telescope and the $HST/ACS$ observation.
They derived a distance modulus of $(m-M)_{0}=24.48\pm0.08\pm0.16$
by comparing the TRGB magnitude of IC 10 to that of the SMC. 
They adopted a reddening of $E(B-V)=0.95\pm0.15$, and their distance modulus 
of IC 10 is 0.21 magnitude larger than that of this study.
If the same reddening of $E(B-V)=0.98$ as in this study is used, 
their distance modulus, $(m-M)_{0} =24.31$, will be very similar to ours. 

Recently \citet{san08} estimated the distance to IC 10 by comparing the
magnitude of the TRGB in the $I$ band to those of the SMC and two Galactic globular clusters,
47 Tuc, and $\omega$ Cen. Adopting a total reddening of IC 10, $E(B-V)=0.78\pm0.06$, and
a distance to the SMC, $(m-M)_{0}=18.75$, they obtained a TRGB distance modulus of IC 10
of $(m-M)_{0}=24.51\pm0.08$. 
If the same reddening of $E(B-V)=0.98$ as in this study is used,
we derive a value $(m-M)_{0}=24.02$. 
If we take a longer distance scale for the SMC of $(m-M)_0=18.93$ \citep{kel06}, 
then their distance
modulus, $(m-M)_{0} =24.20$, is very close to ours.  We note that the reddening value of IC 10 derived by
matching visually the blue sequence of IC 10 to that of blue stars in the SMC
cluster NGC 346 in Figure 3 of \citet{san08} might be a lower limit, because the
reddened RGB sequence of IC 10 is still bluer than that of the SMC stars.

\section{Summary}

We estimated the reddening and distance of the sta burst galaxy IC 10 
using the $JHK_{S}$ photometry obtained from the Subaru/MOIRCS
and $UBV$ photometry of IC 10 given by the Local Group Survey \citep{mas07}.
Primary results are summarized as follows. 
\begin{enumerate}

 \item  We presented $JHK_{S}$ photometry of  $\sim52,000$ stars in the central $4\arcmin\times7\arcmin$ field 
of IC 10 derived from deep images obtained using MOIRCS at the Subaru telescope.
 
 \item  We estimated the foreground reddening of IC 10 using the $UBV$ 
photometry of  foreground MS stars provided by the Local Group Survey \citep{mas07}, 
obtaining $E(B-V)=0.52\pm0.04$.
We also derived a value for the total reddening (including the internal reddening) of $E(B-V)=0.98\pm0.06$,
using the $(U-B)-(B-V)$ diagram of early-type stars in IC 10, and
using a comparison of the RGB loci of IC 10 and the SMC in the $K_S - (J-{K_S})$ CMD.

 \item  We derived relations between the metallicity [Fe/H]$_{CG97}$  
and the slope of the RGB  in the $K_S - (J-K_{S})$ CMD
in several magnitude steps for 20 Galactic globular clusters,
using the 2MASS point source catalog. 
Using these calibrations, we estimated the mean metallicity of the RGB stars in IC 10
to be  [Fe/H]$_{CG97}$ $=-1.08\pm0.28$.

\item The apparent magnitude of the TRGB  is estimated to be 
 $K_{S} = 18.28\pm0.01$.
Then, we derived a distance modulus for IC 10 of
 $(m-M)_{0}=24.27\pm0.03$(random)$\pm0.18$(systematic) 
for a total reddening of $E(B-V)=0.98$, 
corresponding to the distance of 
$d=715^{+10}_{-10}$ $^{+62}_{-57}$ kpc. 
This confirms that IC 10 is a member of the Local Group.

\end{enumerate}

\acknowledgments
This work was 
supported in part by a grant
(R01-2007-000-20336-0) from the Basic Research Program of the
Korea Science and Engineering Foundation. The authors are grateful to the Director 
of the Subaru telescope for allocation of the observing time for this project.

\clearpage

\begin{center}
\begin{deluxetable}{cccc}
\tablecaption{Observation Log for IC 10\label{tbl-observation}}
\tablewidth{0pt}
\tabletypesize{\normalsize}
\tablecolumns{3}\tablehead{
\colhead{Filter}&
\colhead{Exposure Time}&
\colhead{Date}\\
\colhead{(1)}&
\colhead{(2)}&
\colhead{(3)}
}
\startdata
$J$        & 13 s, $9\times100$ s & 2005-12-09\\
$H$        & 13 s, $9\times 75$ s & 2005-12-09\\
$K_{S}$    & 13 s, $9\times135$ s & 2005-12-09\\
\enddata
\tablecomments{
Col. (1): Filter. 
Col. (2): Exposure time. 
Col. (3): Observation date (UT). 
}
\end{deluxetable}
\end{center}

\begin{center}
\begin{deluxetable}{crccr}
\tablecaption{Parameters for Observed Regions\label{tbl-region}}
\tablewidth{0pt}
\tabletypesize{\normalsize}
\tablecolumns{5}\tablehead{
\colhead{Region}&
\colhead{Area}&
\colhead{Number of Stars}&
\colhead{Number Density}&
\colhead{Remark}\\
\colhead{}&
\colhead{[$\arcsec^{2}$]}&
\colhead{[number]}&
\colhead{[number/$\arcsec^{2}$]}&
\colhead{}\\
\colhead{(1)}&
\colhead{(2)}&
\colhead{(3)}&
\colhead{(4)}&
\colhead{(5)}
}
\startdata
RA  & 105564 & $52,054$ & 0.49 & the entire region \\
R1  & 35354  & $12,299$ & 0.35 & the northern outer region \\
R2  & 35190  & $26,004$ & 0.74 & the central region \\
R3  & 35019  & $13,751$ & 0.39 & the southern outer region \\
\enddata
\tablecomments{
Col. (1): Region name. 
RA indicates the entire observed region, and R1, R2,
and R3 indicate three subregions (see Figure \ref{fig-ic10region_dpi150}.).
Col. (2): Geometric area of the selected region in units of arcsec$^{2}$.
Col. (3): Number of detected stars in each region. 
Col. (4): Surface number density of the stars in each region in units of 
number/arcsec$^{2}$.
Col. (5): Remark.
}
\end{deluxetable}
\end{center}

\begin{center}
\begin{deluxetable}{ccccccccccc}
\tablecaption{Parameters for Galactic Globular Clusters\label{tbl-gc}}
\tablewidth{0pt}
\tabletypesize{\scriptsize}
\tablecolumns{6}\tablehead{
\colhead{Name}&
\colhead{Other Name}&
\colhead{[Fe/H]$_{CG97}$}&
\colhead{$S05$}&
\colhead{$S10$}&
\colhead{$S15$}&
\colhead{$S20$}&
\colhead{$S25$}&
\colhead{$S30$}&
\colhead{$S35$}&
\colhead{$S40$}\\
\colhead{(1)}&
\colhead{(2)}&
\colhead{(3)}&
\colhead{(4)}&
\colhead{(5)}&
\colhead{(6)}&
\colhead{(7)}&
\colhead{(8)}&
\colhead{(9)}&
\colhead{(10)}&
\colhead{(11)}
}
\startdata
    NGC 104 &   47Tuc & $   -0.70 $ & $   -5.87 $ & $   -6.12 $ & $   -6.39 $ & $   -6.66 $ & $   -6.94 $ & $   -7.22 $ & $   -7.52 $ & $   -7.82 $  \\
    NGC 288 &         & $   -1.07 $ & $   -5.22 $ & $   -5.65 $ & $   -6.11 $ & $   -6.59 $ & $   -7.10 $ & $   -7.61 $ & $   -8.15 $ & $   -8.69 $  \\
    NGC 362 &         & $   -1.15 $ & $   -7.74 $ & $   -7.98 $ & $   -8.22 $ & $   -8.48 $ & $   -8.73 $ & $   -8.99 $ & $   -9.25 $ & $   -9.52 $  \\
   NGC 1851 &         & $   -1.08 $ & $   -5.12 $ & $   -5.64 $ & $   -6.16 $ & $   -6.67 $ & $   -7.19 $ & $   -7.70 $ & $   -8.21 $ & $   -8.73 $  \\
   NGC 1904 &     M79 & $   -1.37 $ & $   -7.45 $ & $   -8.22 $ & $   -8.98 $ & $   -9.72 $ & $  -10.46 $ & $  -11.20 $ & $  -11.93 $ & $  -12.65 $  \\
   NGC 5466 &         & $   -2.14 $ & $  -18.54 $ & $  -18.72 $ & $  -18.91 $ & $  -19.12 $ & $  -19.34 $ & $  -19.58 $ & $  -19.83 $ & $  -20.08 $  \\
   NGC 6121 &      M4 & $   -1.19 $ & $   -7.80 $ & $   -7.98 $ & $   -8.19 $ & $   -8.40 $ & $   -8.63 $ & $   -8.88 $ & $   -9.15 $ & $   -9.43 $  \\
   NGC 6171 &    M107 & $   -0.87 $ & $   -5.60 $ & $   -6.00 $ & $   -6.43 $ & $   -6.87 $ & $   -7.32 $ & $   -7.79 $ & $   -8.27 $ & $   -8.76 $  \\
   NGC 6205 &     M13 & $   -1.39 $ & $   -7.42 $ & $   -8.00 $ & $   -8.58 $ & $   -9.17 $ & $   -9.76 $ & $  -10.36 $ & $  -10.95 $ & $  -11.55 $  \\
   NGC 6304 &         & $   -0.75 $ & $   -5.11 $ & $   -5.58 $ & $   -6.07 $ & $   -6.56 $ & $   -7.07 $ & $   -7.58 $ & $   -8.10 $ & $   -8.62 $  \\
   NGC 6624 &         & $   -0.63 $ & $   -7.45 $ & $   -7.63 $ & $   -7.83 $ & $   -8.03 $ & $   -8.26 $ & $   -8.50 $ & $   -8.76 $ & $   -9.03 $  \\
   NGC 6637 &     M69 & $   -0.77 $ & $   -7.44 $ & $   -7.60 $ & $   -7.77 $ & $   -7.96 $ & $   -8.16 $ & $   -8.38 $ & $   -8.62 $ & $   -8.88 $  \\
   NGC 6656 &     M22 & $   -1.48 $ & $   -8.83 $ & $   -9.08 $ & $   -9.37 $ & $   -9.69 $ & $  -10.06 $ & $  -10.46 $ & $  -10.91 $ & $  -11.39 $  \\
   NGC 6779 &     M56 & $   -1.75 $ & $   -4.38 $ & $   -6.09 $ & $   -7.53 $ & $   -8.83 $ & $  -10.04 $ & $  -11.20 $ & $  -12.30 $ & $  -13.37 $  \\
   NGC 6809 &     M55 & $   -1.61 $ & $  -13.13 $ & $  -13.47 $ & $  -13.84 $ & $  -14.23 $ & $  -14.64 $ & $  -15.08 $ & $  -15.54 $ & $  -16.01 $  \\
   NGC 6838 &     M71 & $   -0.70 $ & $   -5.88 $ & $   -6.10 $ & $   -6.38 $ & $   -6.69 $ & $   -7.06 $ & $   -7.47 $ & $   -7.92 $ & $   -8.41 $  \\
   NGC 7078 &     M15 & $   -2.12 $ & $  -16.46 $ & $  -17.14 $ & $  -17.83 $ & $  -18.53 $ & $  -19.24 $ & $  -19.97 $ & $  -20.71 $ & $  -21.47 $  \\
   NGC 7089 &      M2 & $   -1.34 $ & $   -7.61 $ & $   -8.47 $ & $   -9.33 $ & $  -10.18 $ & $  -11.03 $ & $  -11.87 $ & $  -12.71 $ & $  -13.55 $  \\
   NGC 7099 &     M30 & $   -1.91 $ & $   -9.12 $ & $  -10.06 $ & $  -10.99 $ & $  -11.88 $ & $  -12.76 $ & $  -13.62 $ & $  -14.47 $ & $  -15.31 $  \\
      Pal8 &         & $   -0.61 $ & $   -7.00 $ & $   -7.16 $ & $   -7.34 $ & $   -7.53 $ & $   -7.74 $ & $   -7.96 $ & $   -8.19 $ & $   -8.44 $   \\
\enddata
\tablecomments{
Col. (1): Globular cluster names.
Col. (2): Other names of the globular clusters.
Col. (3): Mean metallicity of the globular clusters in CG97 from the literature (e.g., \citet{har96};
\citet{fer99}; \citet{fer00}; \citet{val07}).
Col. (4)-Col. (11): S05, S10, S15, S20, S25, S30, S35, and S40 represent the slope of the RGB
in the magnitude range between the TRGB and 0.5, 1.0, 1.5, 2.0, 2.5,
3.0, 3.5, and 4.0 magnitude fainter than that of TRGB in the $K_{S} - (J-K_{S})$ CMD, respectively.
}
\end{deluxetable}
\end{center}

\begin{center}
\begin{deluxetable}{ccccccc}
\tablecaption{TRGB Distance Estimates for IC 10\label{tbl-dm}}
\tablewidth{0pt}
\tabletypesize{\normalsize}
\tablecolumns{6}\tablehead{
\colhead{Filter}&
\colhead{Region}&
\colhead{$S_{15}$}&
\colhead{[Fe/H]$_{CG97}$}&
\colhead{m$_{TRGB}$}&
\colhead{M$_{TRGB}$}&
\colhead{$(m-M)_{0}$}\\
\colhead{(1)}&
\colhead{(2)}&
\colhead{(3)}&
\colhead{(4)}&
\colhead{(5)}&
\colhead{(6)}&
\colhead{(7)}
}
\startdata
J       & RA & $  -7.56 $ & $  -1.08 \pm 0.28 $ & $  19.78 \pm   0.01 $ & $  -5.33 \pm 0.20 $ & $  24.16 \pm   0.06 \pm 0.20 $ \\
J       & R1 & $  -6.92 $ & $  -1.01 \pm 0.28 $ & $  19.76 \pm   0.01 $ & $  -5.36 \pm 0.20 $ & $  24.17 \pm   0.06 \pm 0.20 $ \\
J       & R2 & $  -7.34 $ & $  -1.06 \pm 0.28 $ & $  19.76 \pm   0.01 $ & $  -5.34 \pm 0.20 $ & $  24.15 \pm   0.06 \pm 0.20 $ \\
J       & R3 & $  -7.52 $ & $  -1.08 \pm 0.28 $ & $  19.84 \pm   0.01 $ & $  -5.34 \pm 0.20 $ & $  24.22 \pm   0.06 \pm 0.20 $ \\
\hline
H       & RA & $  -7.56 $ & $  -1.08 \pm 0.28 $ & $  18.73 \pm   0.01 $ & $  -6.20 \pm 0.16 $ & $  24.33 \pm   0.04 \pm 0.16 $ \\
H       & R1 & $  -6.92 $ & $  -1.01 \pm 0.28 $ & $  18.75 \pm   0.01 $ & $  -6.23 \pm 0.16 $ & $  24.38 \pm   0.04 \pm 0.16 $ \\
H       & R2 & $  -7.34 $ & $  -1.06 \pm 0.28 $ & $  18.68 \pm   0.02 $ & $  -6.21 \pm 0.16 $ & $  24.28 \pm   0.04 \pm 0.16 $ \\
H       & R3 & $  -7.52 $ & $  -1.08 \pm 0.28 $ & $  18.72 \pm   0.04 $ & $  -6.20 \pm 0.16 $ & $  24.32 \pm   0.05 \pm 0.16 $ \\
\hline
K$_{S}$ & RA & $  -7.56 $ & $  -1.08 \pm 0.28 $ & $  18.28 \pm   0.01 $ & $  -6.35 \pm 0.18 $ & $  24.26 \pm   0.03 \pm 0.18 $ \\
K$_{S}$ & R1 & $  -6.92 $ & $  -1.01 \pm 0.28 $ & $  18.29 \pm   0.01 $ & $  -6.39 \pm 0.18 $ & $  24.29 \pm   0.03 \pm 0.18 $ \\
K$_{S}$ & R2 & $  -7.34 $ & $  -1.06 \pm 0.28 $ & $  18.30 \pm   0.02 $ & $  -6.37 \pm 0.18 $ & $  24.30 \pm   0.03 \pm 0.18 $ \\
K$_{S}$ & R3 & $  -7.52 $ & $  -1.08 \pm 0.28 $ & $  18.26 \pm   0.01 $ & $  -6.36 \pm 0.18 $ & $  24.25 \pm   0.03 \pm 0.18 $ \\
\enddata
\tablecomments{
Col. (1): Filter. 
Col. (2): Region in the observed field. (see Figure \ref{fig-ic10region_dpi150}).
Col. (3): Slope of the RGB in the magnitude range between the TRGB and
1.5 magnitude fainter than that of the TRGB in the $K_{S}-(J-K_{S})$ CMD.
Col. (4): Metallicity estimated with equation (3) using the RGB slope
in the $K_{S} - (J-K_{S}) $ CMD.
Col. (5): Apparent magnitude of the TRGB.
Col. (6): Absolute magnitude of the TRGB derived with equations
(7)-(9) in \citet{val04}. 
Col. (7): Distance modulus of IC 10. The quoted errors consist of
(random error)$\pm$(systematical error). The random error comes from TRGB
detection uncertainty plus reddening uncertainty. The systematical
error is due to the uncertainty in the calibration equation of the
absolute magnitude of the TRGB.
}
\end{deluxetable}
\end{center}

\begin{center}
\begin{deluxetable}{ccccr}
\tablecaption{A Summary of Distance Estimates for IC 10\label{tbl-other}}
\tablewidth{0pt}
\tabletypesize{\small}
\tablecolumns{5}\tablehead{
\colhead{Method}&
\colhead{$(m-M)_{0}$}&
\colhead{$E(B-V)$}&
\colhead{Filter}&
\colhead{Reference}\\
\colhead{(1)}&
\colhead{(2)}&
\colhead{(3)}&
\colhead{(4)}&
\colhead{(5)}
}
\startdata
WR/Blue plume    & $24.90$            & $0.75-0.80$   & sp/{\it{BV}}       & 1\\
Cepheid          & $24.59\pm0.30$     & $0.94$        & {\it{gri}}         & 2\\
Cepheid          & $24.57\pm0.21$     &               & {\it{JHK}}         & 3\\
Cepheid          & $24.10\pm0.20$     & $1.16\pm0.08$ & {\it{VI}}          & 4\\
TRGB             & $23.51\pm0.19$     & $1.16\pm0.08$ & {\it{VI}}          & 4\\
Red Super Giants & $23.86\pm0.12$     & $1.05\pm0.10$ & {\it{JHK}}         & 5\\
TRGB             & $24.95\pm0.20$     & 0.77          & {\it{F555W,F814W}} & 6\\ 
Carbon Star      & $24.35\pm0.11$     & variable      & {\it{R,I,CN,TiO}}  & 7\\ 
TRGB             & $24.48\pm0.08$     & 0.95          & {\it{F814W, K$'$}} & 8\\ 
TRGB             & $24.51\pm0.08$     & $0.78\pm0.06$ & {\it{F555W,F814W}} & 9\\
\hline
TRGB             & $24.27\pm0.03\pm0.18$     & $0.98\pm0.06$ & {\it{$JHK_{S}$}}        & 10\\
\enddata
\tablecomments{
Col. (1): Methods for determining the distance to IC 10.
Col. (2): Distance modulus.
Col. (3): Adopted reddening of IC 10.
Col. (4): Filters.
Col. (5): Reference.
1:\citet{mas95}, 2:\citet{sah96}, 3:\citet{wil96},  4:\citet{sak99}, 
5:\citet{bor00}, 6:\citet{hun01}, 7:\citet{dem04}, 8:\citet{vac07}, 
9:\citet{san08},  and 10:This study.
}
\end{deluxetable}
\end{center}

\clearpage

\epsscale{0.6}\plotone{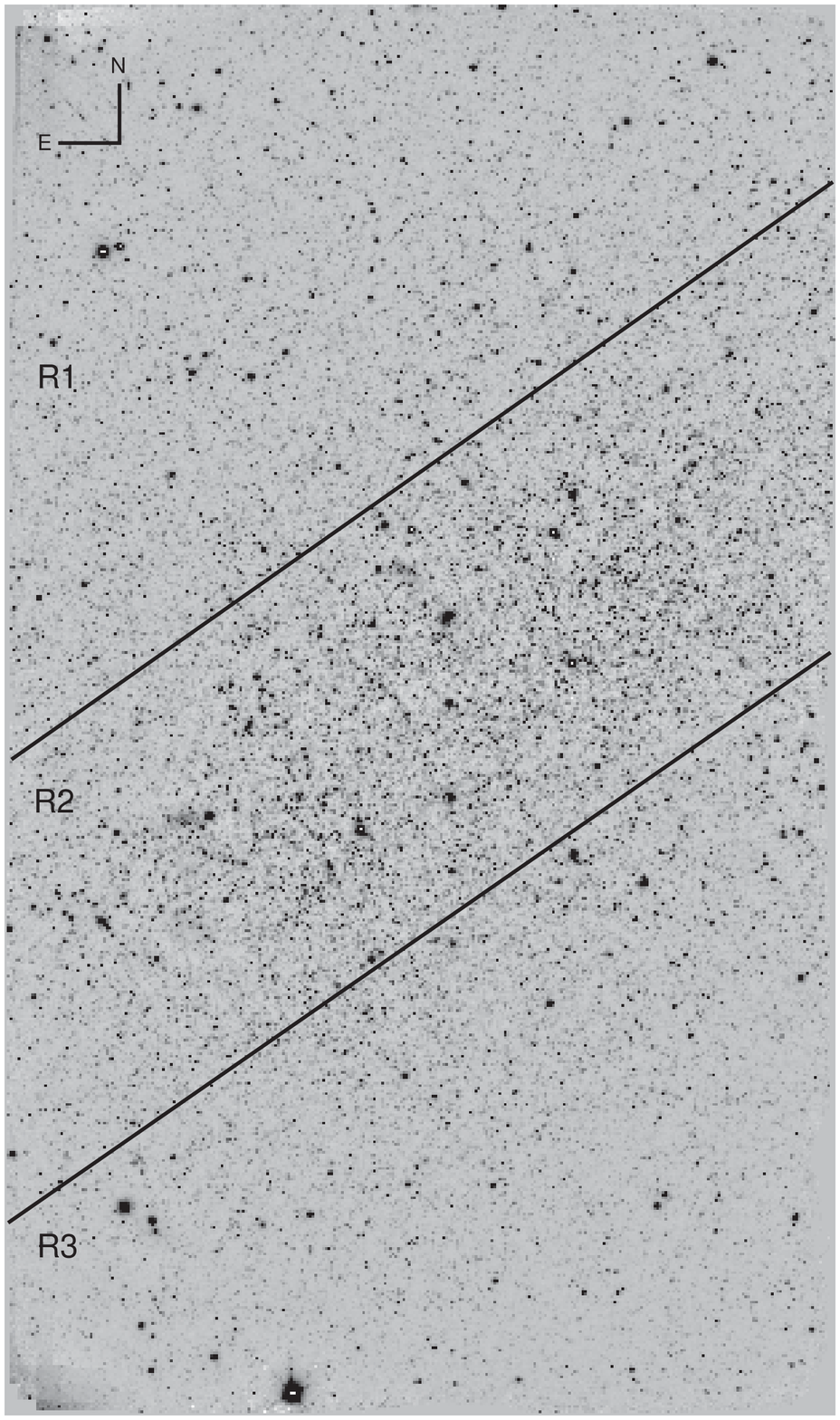} 
\figcaption[ic10region_dpi150.ps]{
A gray scale map of $K_{S}$ band image of IC 10. North is up and east to the left.
The size of the field of the view is $4\arcmin \times 7\arcmin$.
The observed field is divided into three subregions (R1, R2, and R3)
for the analysis.
\label{fig-ic10region_dpi150}}
\clearpage

\epsscale{1.0}\plotone{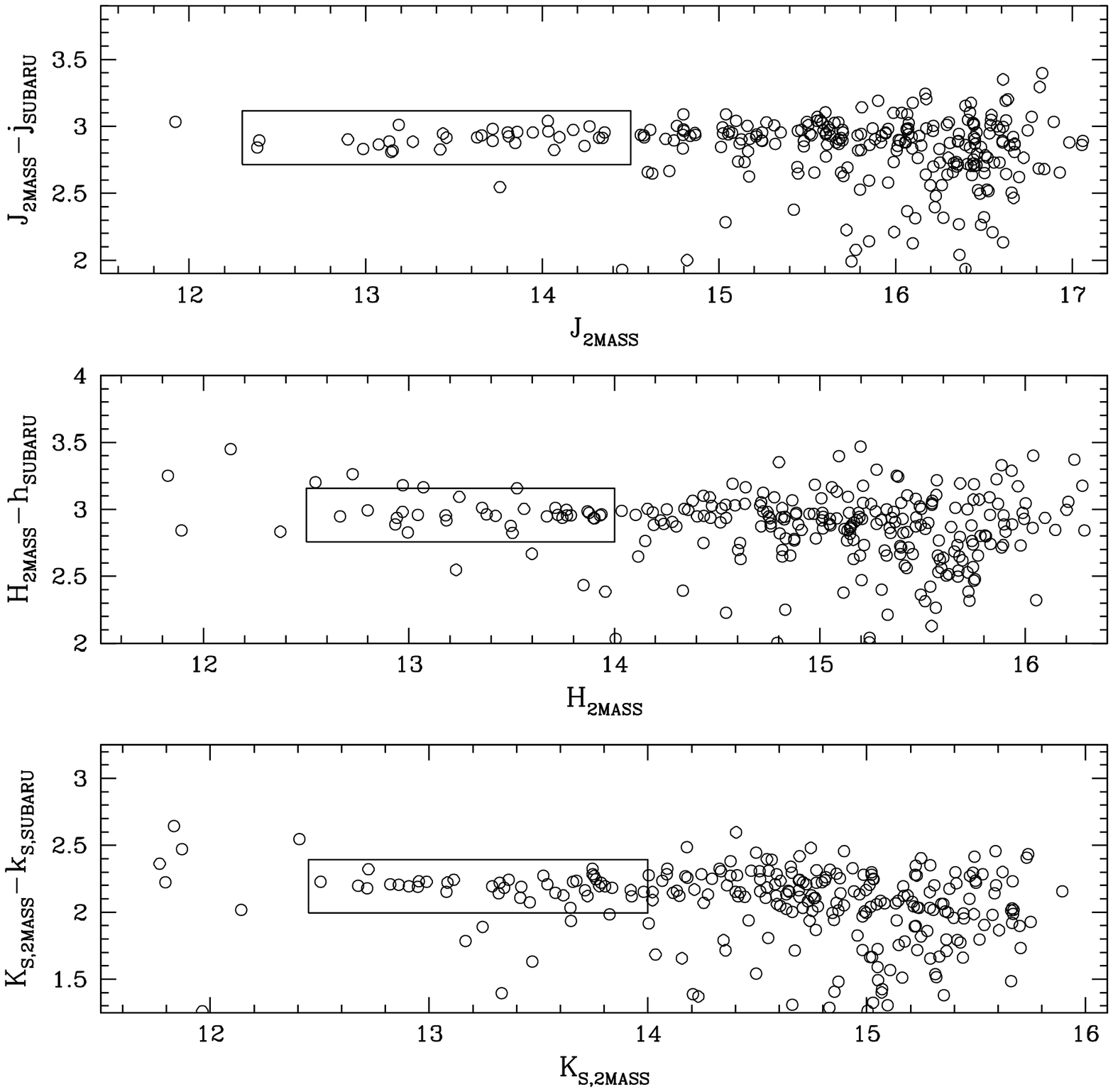} 
\figcaption[calib.ps]{
Differences between the 2MASS magnitudes (upper cases) and the instrumental magnitudes 
from the present SUBARU observations (lower cases) for the stars common between the 2MASS
point source catalog and this study.  
The data points inside the box in each panel are used for standard calibration. 
\label{fig-calib}}
\clearpage

\plotone{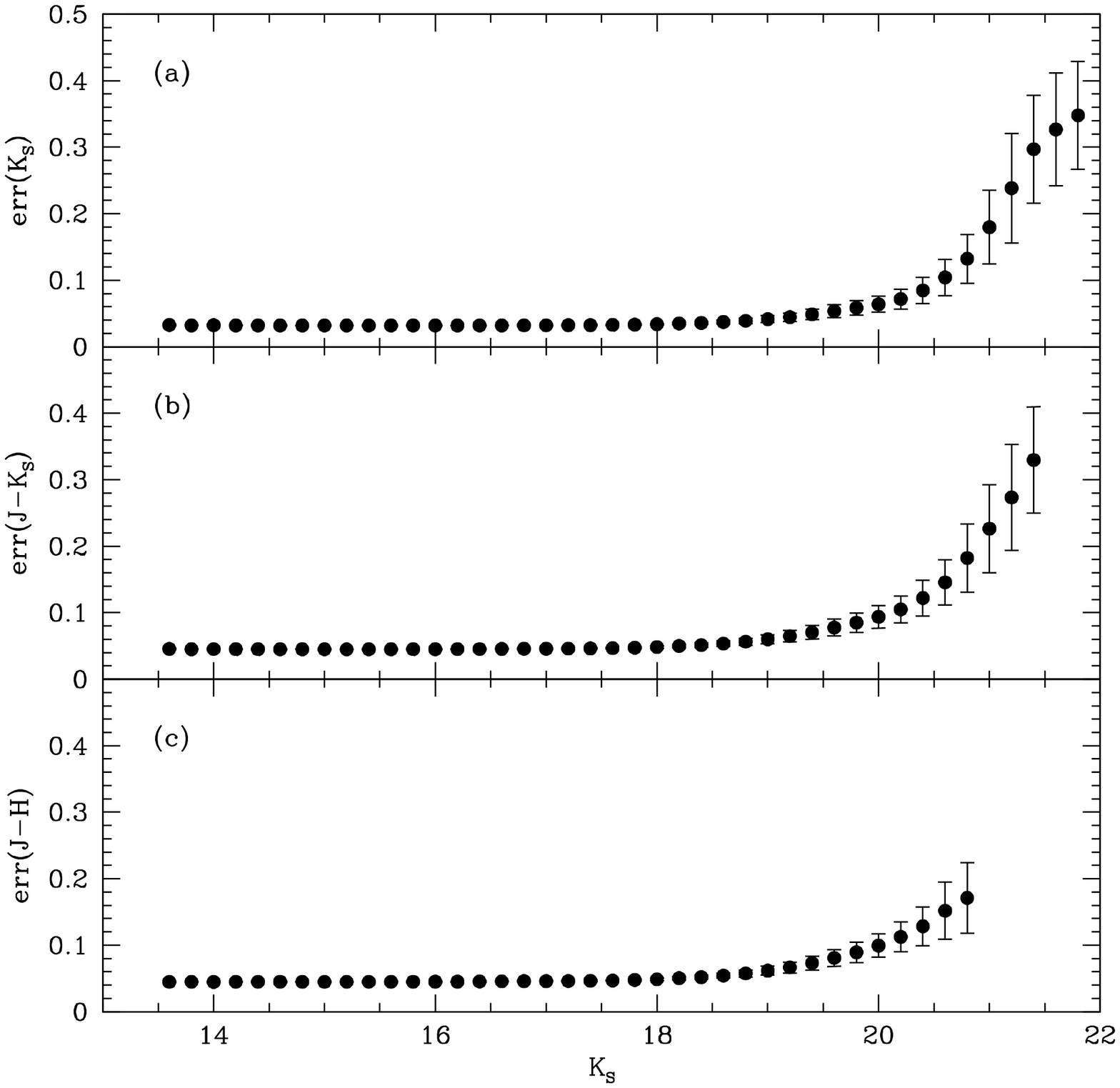} 
\figcaption[mag_err.ps]{
Mean photometric errors in the $K_{S}$ magnitude and the $(J-K_{S})$ and $(J-H)$ colors
as a function of the $K_{S}$ magnitude.
The error bar indicates $1\sigma$ error in each magnitude bin.
\label{fig-mag_err}}
\clearpage

\plotone{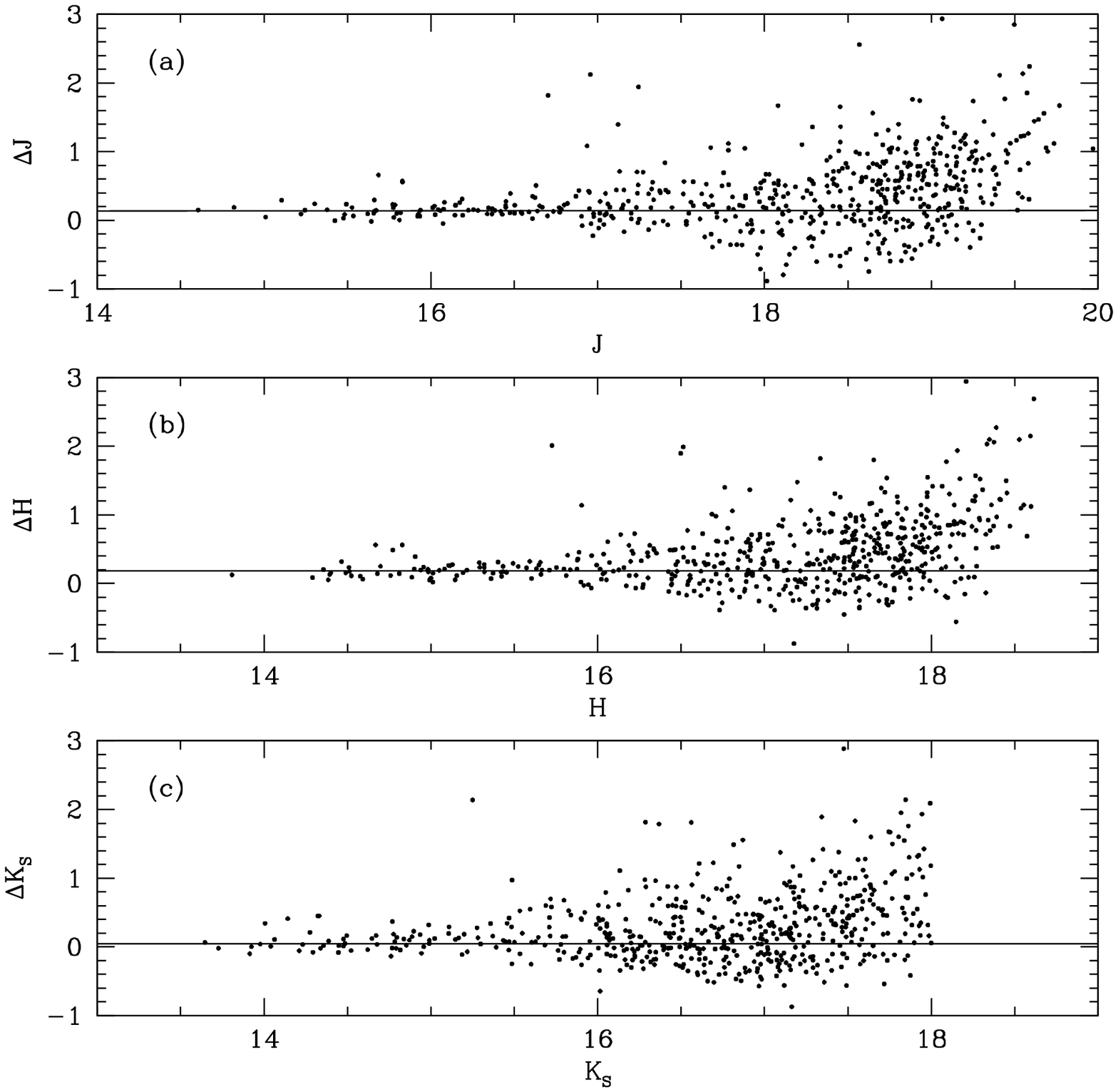} 
\figcaption[comp_bori.ps]{
A comparison of the magnitudes in this study and those in \cite{bor00} for
(a) $J$, (b) $H$, and (c) $K_{S}$ bands, respectively. 
In each panel, the solid line represents a mean difference with 
$2\sigma$ clipping for bright stars with $J<17$, $H<16$, and $K_{S}<16$ mag : 
$\Delta J=0.14$ mag, $\Delta H=0.18$ mag, and  $\Delta K_{S}=0.05$ mag,
where $\Delta$ means this study minus \cite{bor00}.
\label{fig-comp_bori}}
\clearpage

\plotone{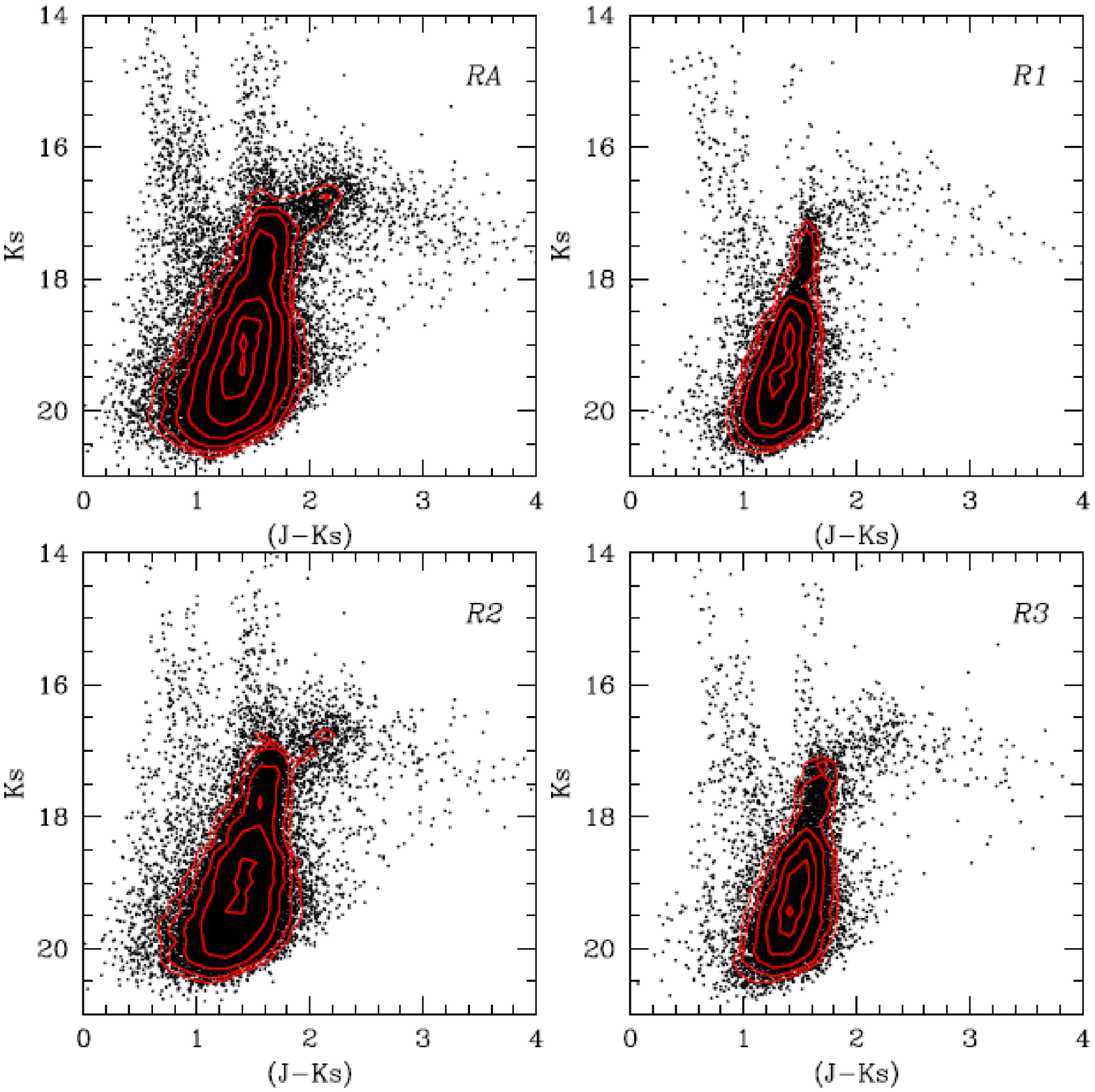} 
\figcaption[cmd_jkk_con_n3.ps]{
$K_{S} - (J-K_{S})$ CMDs of IC 10 for the entire region RA
and three subregions R1, R2, and R3, respectively. 
The number density contour maps are overlayed with solid lines to show 
clearly the morphology of the RGB.
\label{fig-cmd_jkk_con_n3}}
\clearpage

\plotone{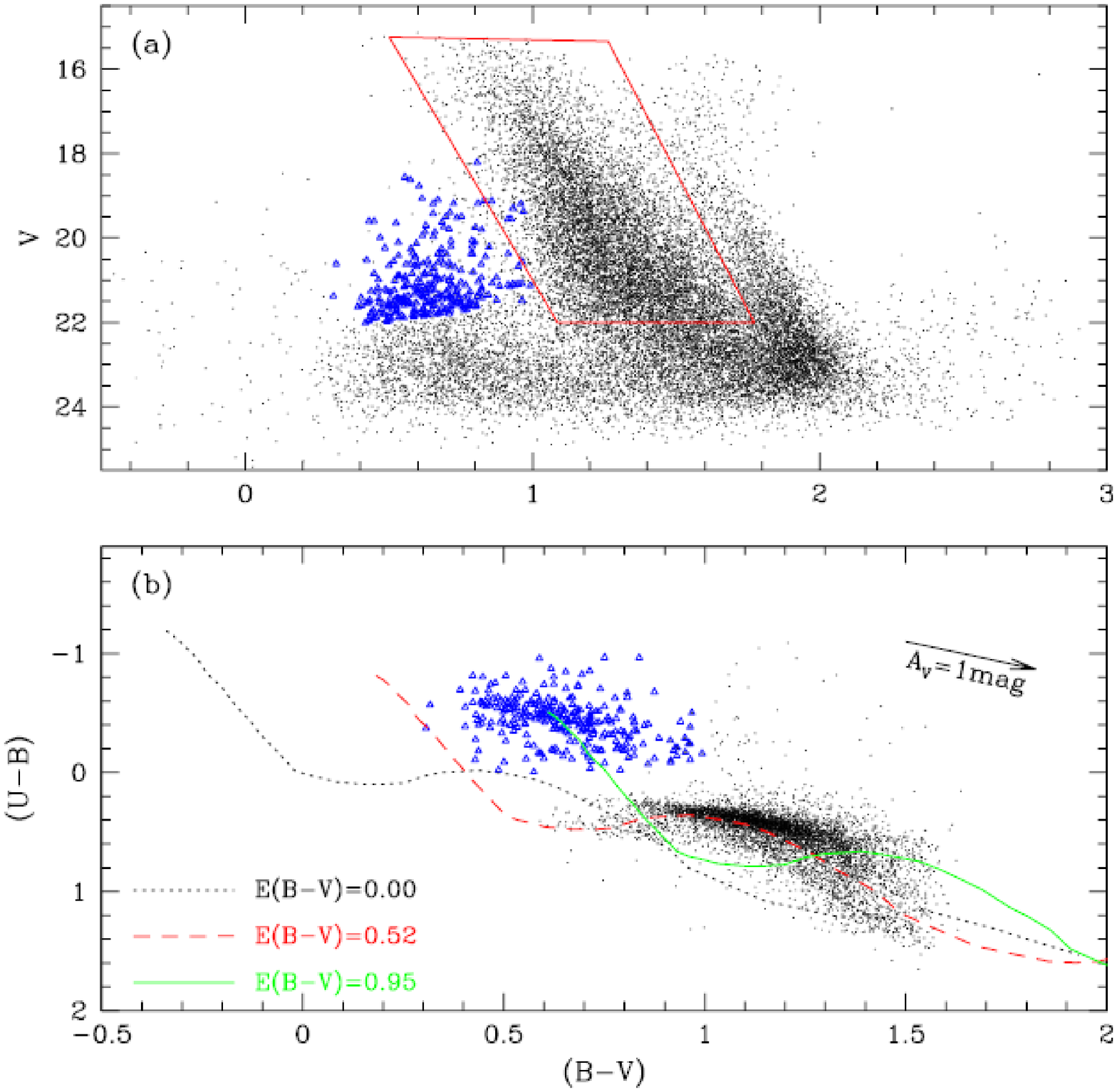} 
\vspace{-10mm}
\figcaption[plot_ic10_ext4paper1.ps]{
(a) The $V-(B-V)$ CMD of the 550 square arcmin 
field of IC 10 obtained from the Local Group
Survey \citep{mas07}. 
Open triangles represent the bright MS stars in IC 10.
The stars inside the tilted box are foreground MS stars 
belonging to the Galaxy.  
(b) The $(U-B)-(B-V)$ color-color diagram of IC 10.
Open triangles and dots represent the MS stars
of IC 10 and the Galaxy, respectively, as selected from (a). 
The dotted, long dashed, and solid lines represent
the intrinsic MS sequence \citep{sch82} reddened according to $E(B-V)=0.00$, 0.52, and 0.95, respectively.
The arrow indicates the reddening direction 
and its length corresponds to one magnitude extinction for $V$ band.
\label{fig-plot_ic10_ext4paper1}}
\clearpage

\plotone{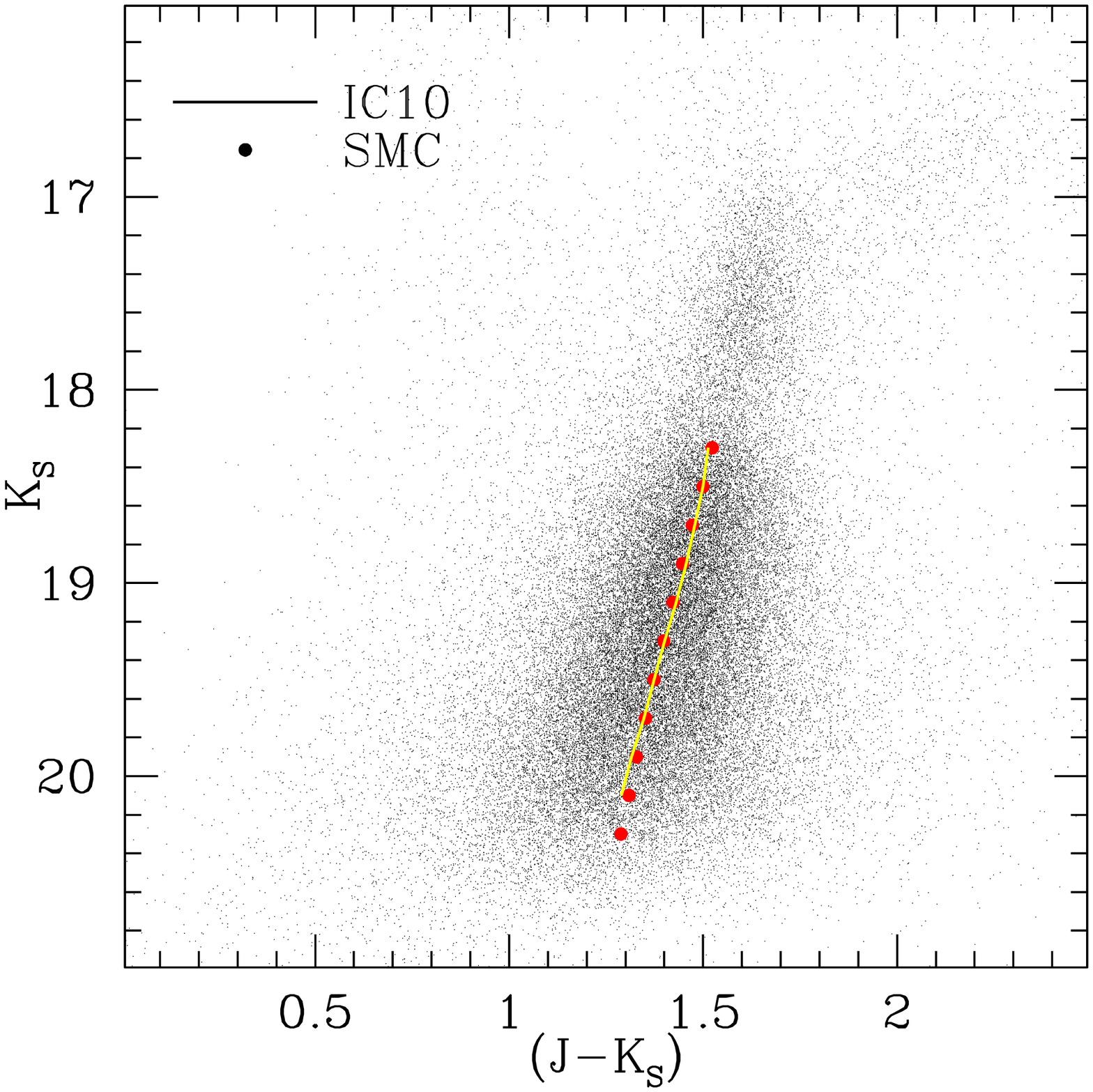} 
\figcaption[plot_ic10_ext4paper2.ps]{
The mean locus of the RGB in IC 10 ({\it{solid line}}) 
in the $K_{S} - (J-K_{S})$ CMD 
in comparison with that of the SMC ({\it{filled circle}}) derived using the data given by \citet{kat07}.
The locus of the SMC was shifted according to the distance and reddening of IC 10.
\label{fig-plot_ic10_ext4paper2}}
\clearpage

\plotone{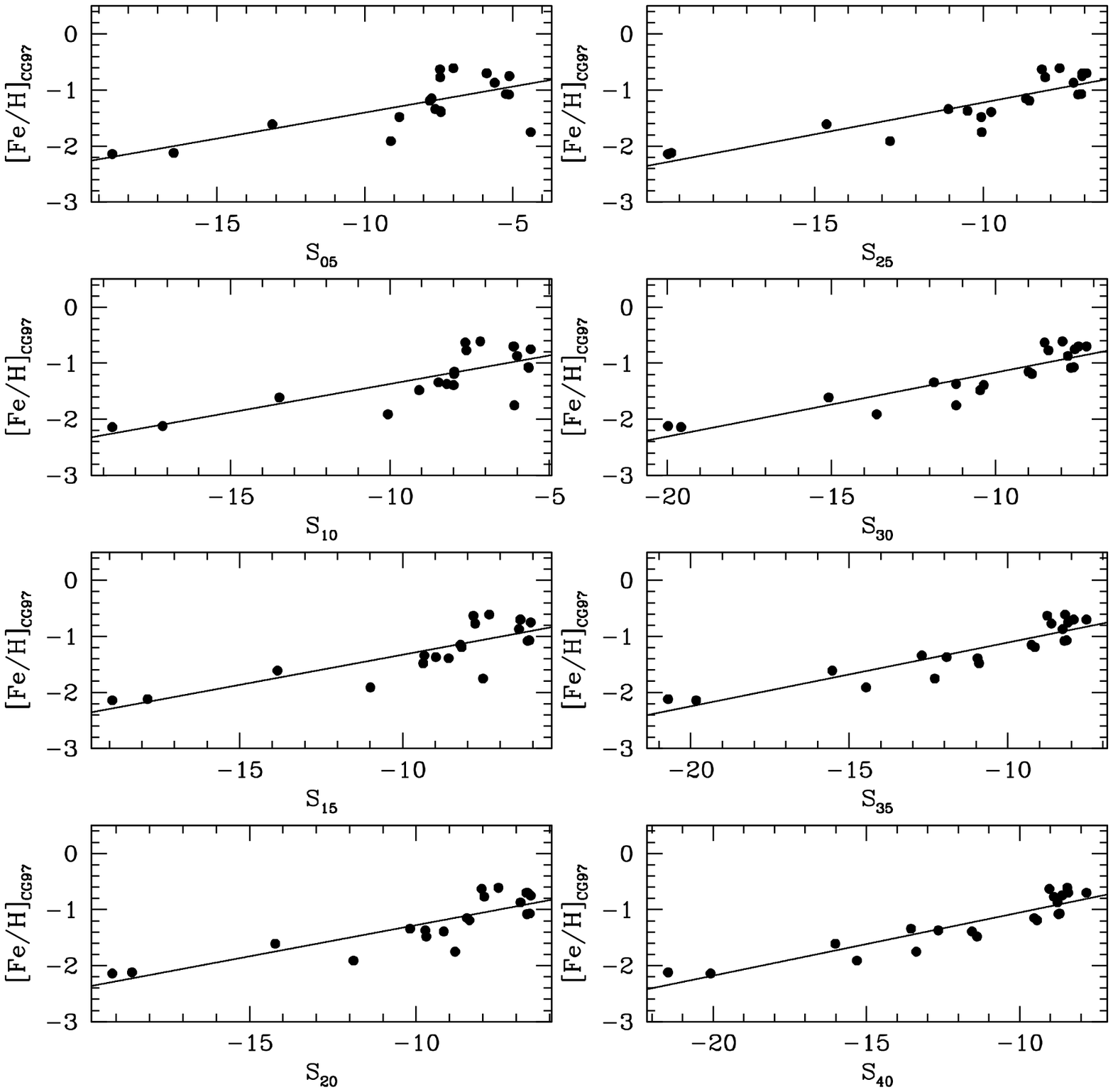} 
\figcaption[slopefeohfit.ps]{
Relation between the RGB slope and [Fe/H]$_{CG97}$
in the $K_{S} - (J-K_{S})$ CMD.
$S_{05}$, $S_{10}$, $S_{15}$, $S_{20}$, $S_{25}$, $S_{30}$, $S_{35}$, and $S_{40}$ represent the slope of the RGB
in the magnitude range between 
the TRGB and 0.5, 1.0, 1.5, 2.0, 2.5,
3.0, 3.5, and 4.0 magnitude fainter than that of TRGB, respectively. 
The solid line represents the best linear fit to the data.
\label{fig-slopefeohfit}}
\clearpage

\plotone{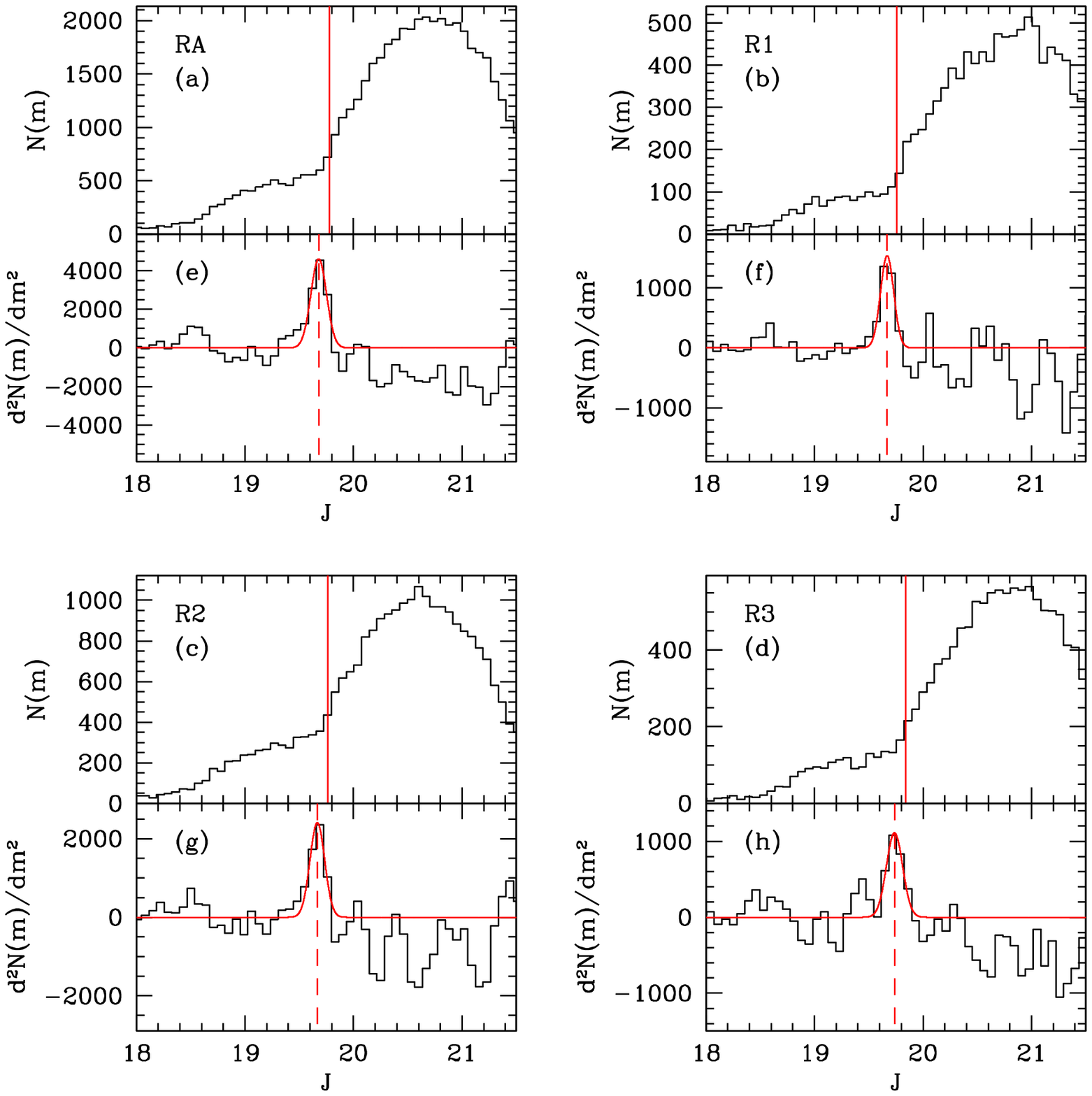} 
\vspace{-20mm}
\figcaption[savgol_fiter_j.ps]{
(a)-(d). The luminosity functions of the stars in the entire region (RA) and three subregions,
(R1, R2, and R3), respectively, 
as a function of the $J$ band magnitude.
The vertical solid line indicates the peak of the Gaussian fit corresponding to the magnitude
of the TRGB after applying the correction factor $\sigma_{2g}$.
(e)-(h). The second derivative of the luminosity functions derived with a Salvizky-Golay filter
in each region. 
The curved solid line indicates the best Gaussian fit centered at the first primary peak of
the second derivative of the luminosity function. 
The vertical dashed line indicates the peak of the Gaussian fit corresponding to the magnitude
of the TRGB before applying the correction factor $\sigma_{2g}$.
\label{fig-savgol_fiter_j}}
\clearpage

\plotone{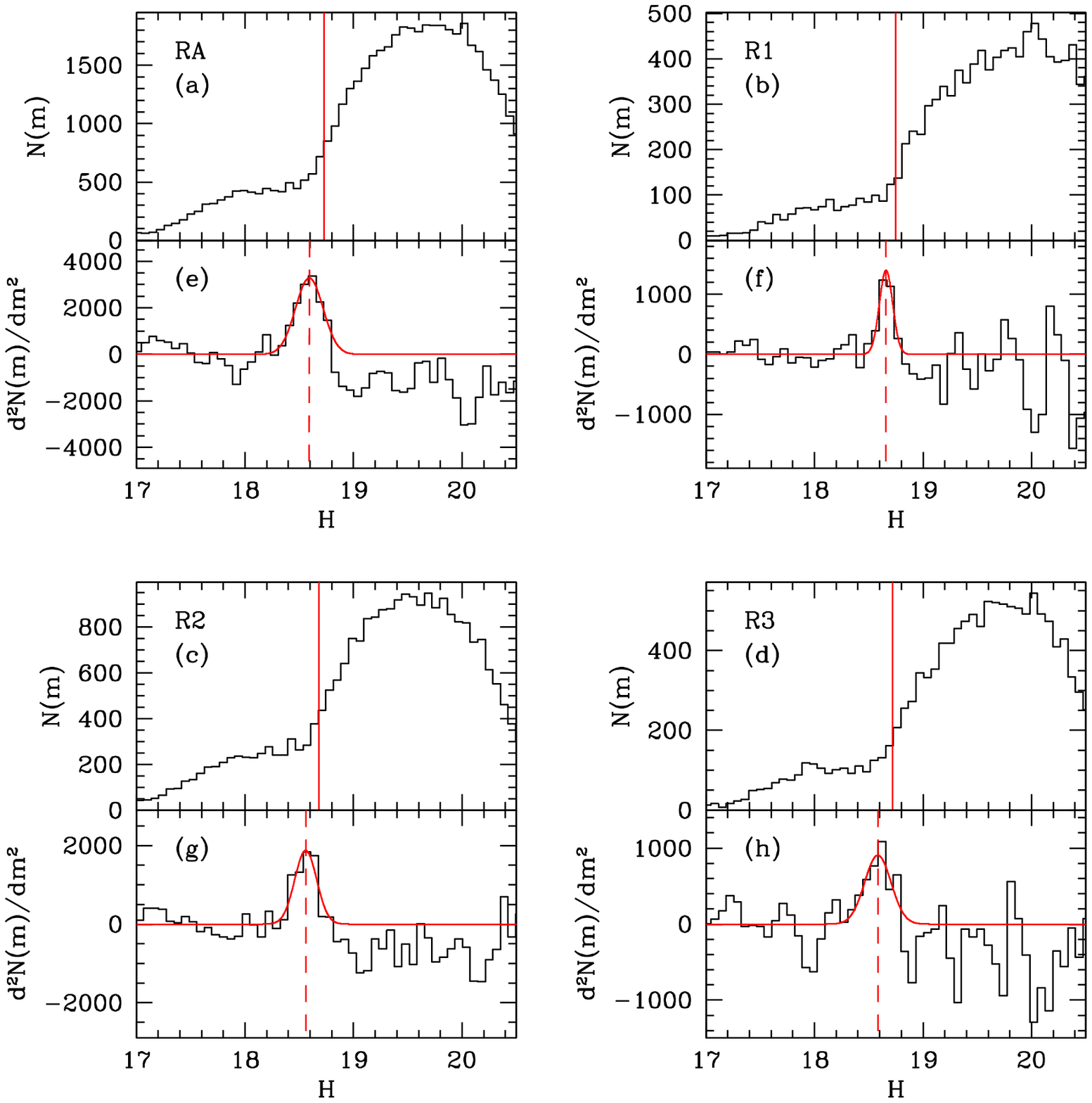} 
\figcaption[savgol_fiter_h.ps]{
Same as Figure \ref{fig-savgol_fiter_j}, but for the $H$ band.
\label{fig-savgol_fiter_h}}
\clearpage

\plotone{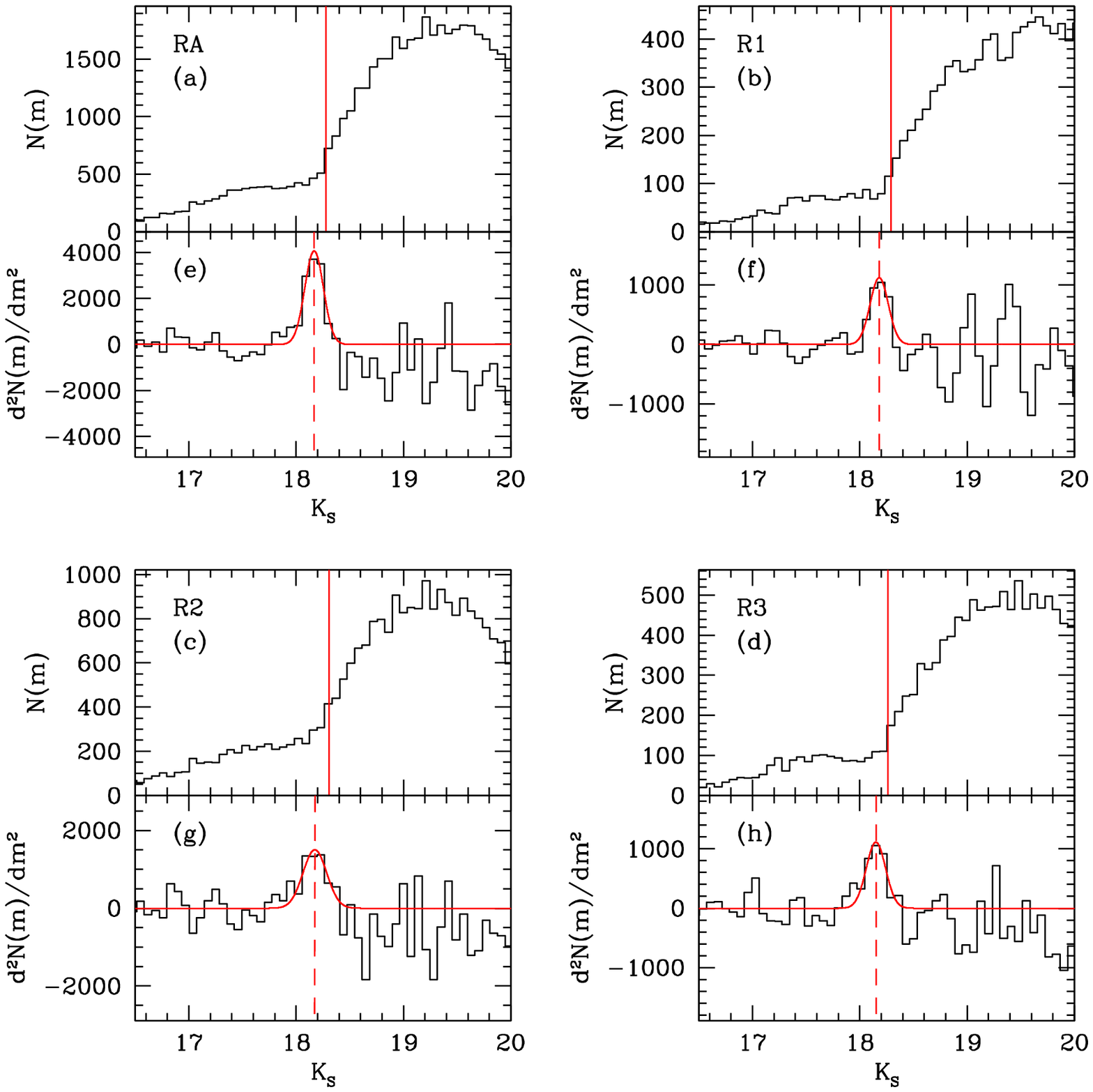} 
\figcaption[savgol_fiter_ks.ps]{
Same as Figure \ref{fig-savgol_fiter_j}, but for the $K_{S}$ band.
\label{fig-savgol_fiter_ks}}
\clearpage

\end{document}